\documentclass[twocolumn,onecolappendix]{aastex62}

\usepackage{float}
\usepackage{textcomp}
\usepackage{amsmath}
\usepackage{mathtools}
\usepackage{natbib}
\usepackage{booktabs}
\usepackage{dcolumn}

\newcommand{\MbhT}{$M_{\rm BH}-kT$ }
\newcommand{\MbhM}{$M_{\rm BH}-M_{\rm bulge}$ }
\newcommand{\MbM}{$M_{\rm BH}-M_{\rm 500}$ }
\newcommand{\MbT}{$M_{\rm bulge} -kT$ }
\newcommand{\neo}{$n_{\rm e,0}$ }
\newcommand{\Ko}{$K_{\rm 0}$ }
\newcommand{\csb}{$C_{\rm sb}$ }
\newcommand{\plm}{$\pm$ }

\shorttitle{Correlating BH mass and the total mass of clusters}
\shortauthors{Phipps et al.}
\begin{document}

\title{\LARGE Expanding the Sample: The Relationship Between the Black Hole Mass of BCGs and the Total Mass of Galaxy Clusters }

\author{Frederika Phipps}
\affil{School of Physics and Astronomy, University of Southampton, Southampton, SO17 1BJ, United Kingdom}
\affiliation{Harvard-Smithsonian Center for Astrophysics, 60 Garden Street, Cambridge, MA 02138, USA}

\author{{\'A}kos Bogd{\'a}n}
\affiliation{Harvard-Smithsonian Center for Astrophysics, 60 Garden Street, Cambridge, MA 02138, USA}

\author{Lorenzo Lovisari}
\affiliation{Harvard-Smithsonian Center for Astrophysics, 60 Garden Street, Cambridge, MA 02138, USA}

\author{Orsolya E. Kov{\'a}cs}
\affiliation{Harvard-Smithsonian Center for Astrophysics, 60 Garden Street, Cambridge, MA 02138, USA}
\affiliation{Konkoly Observatory, MTA CSFK, H-1121 Budapest, Konkoly Thege M. {\'u}t 15-17, Hungary}
\affiliation{E{\"o}tv{\"o}s University, Department of Astronomy, Pf. 32, 1518, Budapest, Hungary}

\author{Marta Volonteri}
\affiliation{Institut d{\'A}strophysique de Paris, Sorbonne Universit{\'e}s, UPMC Univ Paris 6 et CNRS, UMR 7095, 98 bis bd Arago, F-75014 Paris, France}

\author{Yohan Dubois}
\affiliation{Institut d{\'A}strophysique de Paris, Sorbonne Universit{\'e}s, UPMC Univ Paris 6 et CNRS, UMR 7095, 98 bis bd Arago, F-75014 Paris, France}
\begin{abstract}

Supermassive Black Holes (BHs) residing in brightest cluster galaxies (BCGs) are overly massive when considering the local relationships between the BH mass and stellar bulge mass or velocity dispersion. Due to the location of these BHs within the cluster, large-scale cluster processes may aid the growth of BHs in BCGs. In this work, we study a sample of 71 galaxy clusters to explore the relationship between the BH mass, stellar bulge mass of the BCG, and the total gravitating mass of the host clusters. Due to difficulties in obtaining dynamically measured BH masses in distant galaxies, we use the Fundamental Plane relationship of BHs to infer their masses. We utilize X-ray observations taken by \textit{Chandra} to measure the temperature of the intra-cluster medium (ICM), which is a proxy for the total mass of the cluster. We analyze the \MbhT and \MbhM relationships and establish the best- fitting power laws:$\log_{10}(M_{\rm BH} /10^9 M_{\odot})=-0.35+2.08 \log_{10}(kT / 1 \rm keV)$ and $\log_{10}(\rm M_{BH}/10^9M_{\odot})= -1.09+ 1.92 \log_{10}(M_{\rm bulge}/10^{11}M_{\odot})$. Both relations are comparable with that established earlier for a sample of brightest group/cluster galaxies with dynamically measured BH masses. Although both the \MbhT and the \MbhM relationships exhibit large intrinsic scatter, based on Monte Carlo simulations we conclude that dominant fraction of the scatter originates from the Fundamental Plane relationship. We split the sample into cool core and non-cool core resembling clusters, but do not find statistically significant differences in the  \MbhT relation. We speculate that the overly massive BHs in BCGs may be due to frequent mergers and cool gas inflows onto the cluster center.

\vspace{10mm}

\end{abstract}


\section{Introduction} \label{sec:intro}

Throughout the evolution of galaxies, they undergo diverse physical processes, which produce the observed galaxy populations and result in various relations between different galaxy properties \citep{crot06,faber07}. BHs are believed to have a profound effect on the evolution of their host galaxy due to their energetic feedback \citep{rich98,calt09,cheung16}. By precisely measuring the mass of BHs, relationships between the BH mass and the properties of the host galaxies have been established, the most well-known being the correlation between central stellar velocity dispersion ($\sigma$) and BH mass (e.g. \citealt{ferr00,geb00,mccon13,degraf15}), as well as the relationship between stellar bulge mass and BH mass (e.g. \citealt{mag98,har04}). These results have contributed to the development of the current theoretical paradigm, in which the BH and the host galaxies co-evolve and regulate each others growth \citep{fab99,king03,dimat05,hop06,shank06,somer08}. 

The study of BHs and their host galaxies has been extensive, despite the difficulties in measuring the BH masses accurately. However, large-scale structures, in which most galaxies are embedded  \citep{sepp13}, could influence the evolution of BHs. Most galaxy groups/clusters contain a unique type of elliptical galaxy, known as the Brightest Group/Cluster Galaxies (BGGs/BCGs) at their center \citep{craw99,bern07}. Typically, BGGs/BCGs are both the most massive and luminous galaxies in the group/cluster. As these galaxies are located at the bottom of the potential well for these large-scale structures, it is feasible that the BHs of BCGs undergo a different evolution than BHs residing in field or satellite galaxies. Studies of BHs in BCGs pointed out that many of these BHs are over-massive in comparison to the stellar bulge mass or velocity dispersion of the BCG \citep{mccon11,mccon12,mezcua17}. This hints that the large-scale potential of clusters may aid the growth of these BHs.

There is a vast difference between the scales of BHs and galaxy groups/clusters. While the sphere of influence of BHs is $\sim 100 \ \rm{pc}$ for BHs with masses $\sim 10^9M_{\rm \odot}$, galaxy groups/clusters extend to Mpc scales. Therefore, it is appealing to probe if the growth of BHs in BCGs may be influenced by the large-scale structures. From theoretical considerations, a correlation between BH mass and cluster halo temperature is expected \citep{gasp17}. This relation had been previously quantified by \citet{mitt09}, who used the luminosity of the BCG as a proxy for the BH mass. More recently, \citet{bogdan17} investigated using a sample of 17 galaxy groups/clusters, which had dynamically measured BHs in their BGGs/BCGs. By analyzing \textit{XMM-Newton} X-ray observations, they found a tight correlation between the BH and total mass of the group/cluster, which was traced via the gas temperature of the ICM. This relation had an intrinsic scatter in the x and y-axes of $0.22$ and $0.38$. They concluded that the \MbhT relation is tighter and has less scatter than the \MbhM relation (which had scatter of $0.35$ and $0.68$ in each axes), hinting that the BH mass of BGGs/BCGs may be determined by physical processes that are governed by the properties of the large-scale potential. The results of \citet{bogdan17} have been succesfully reproduced in simulations \citep{bass19}. However, the sample of \citet{bogdan17} was relatively small, as it was limited by the available dynamical BH mass measurements in BGGs/BCGS. Whilst this means their masses are more accurate, this also meant that they were unable to investigate these relationships for a notable sample of massive clusters. To further probe the findings of \citet{bogdan17}, it is necessary to extend the sample of galaxy groups and clusters, especially including massive systems.

As measuring the mass of BHs using dynamical methods is challenging for less massive and/or distant BHs,  we must rely on tracers. In this work, we utilize the Fundamental Plane relationship, which defines a plane between BH mass and its luminosity in the X-ray and radio \citep{merloni03,falcke04,kord06,gult09}. The Fundamental Plane relationship for BCGs was investigated by \citet{hlav12} for a sample of 18 BCGs. This work was extended by \citet{mezcua17}, who measured the X-ray and radio properties for a large sample of BHs in BCGs using \textit{Chandra} and \textit{VLBI} data. Here, we use their results to infer both BH masses and bulge masses of the BCGs. While they investigated the Fundamental Plane relationship of both \citet{merloni03} and \citet{plot12}, in this work we opt to use the ``standard" relation of \citet{merloni03} as this relation covers the largest range of radio luminosities and BH masses.  

To derive the total mass of the galaxy clusters, we utilize \textit{Chandra} X-ray observations. Specifically, the temperature of the ICM is a good proxy for the cluster's total mass \citep{horn99,ett13}. Not only is it a good proxy, but through using this method we maintain a straight forward comparison with the work of \citet{bogdan17}. We then use the temperature measured from these X-ray observations to study the BH mass -- cluster temperature relationship. 

The structure of the paper is as follows: in Section 2 we discuss the sample selection process and the observations used to obtain  them. Section 3 describes the analysis of the \textit{Chandra} data. Results are presented in Section 4 and in Section 5 we discuss the implications of these results. We summarize our results in Section 6. 
In this paper, we take the Hubble constant, $H_0$, to be $70 \ \rm{km \ s^{-1} \ {Mpc}^{-1}}$ and $\Omega_{M} = 0.3$ and $\Omega_{\Lambda} = 0.7$. All error bars represent $1\sigma$ uncertainties, unless otherwise mentioned. 

\startlongtable
\begin{deluxetable*}{ccccccc}
\centering
\vspace{-0.5em}
\tablecaption{Analyzed \textit{Chandra} observations in this work.}
\startdata
Name & Obsid & Instrument & redshift & $t_{\rm total}$ & $t_{\rm clean}$ & Date\\
 &  & &  & (ks) & (ks) &  \\  \hline 
A1204 & 2205 & ACIS-I & 0.171 & 23.6 & 20.3 & 2001 Jun 01 \\
A1367 & 514 & ACIS-S &  0.022 & 40.5 & 31.1 & 2000 Feb 26 \\
 & 17201 &ACIS-I & 0.022  &  61.3 & 48.5 & 2016 Jan 31 \\
A1446 & 4975 & ACIS-S & 0.103 & 58.4 & 49.3 & 2004 Sep 29 \\
A1644 & 7922 & ACIS-I & 0.048 &  51.5 & 40.1 & 2007 May 24 \\
A1664 & 7901 & ACIS-S & 0.128 & 36.6 & 30.0 & 2006 Dec 04 \\
A168 & 3203 & ACIS-I & 0.044 & 40.6 & 31.3 & 2002 Feb 05 \\
A1763 & 3591 & ACIS-I & 0.228 & 19.6 & 17.0 & 2003 Oct 28 \\
A1795 & 493 & ACIS-S & 0.063 & 19.6 & 14.4 & 2000 Mar 21 \\
A1930 & 11733 & ACIS-S & 0.132 & 34.5 & 25.9 & 2010 Sep 09 \\
A2009 & 10438 & ACIS-I & 0.153 & 19.9 & 8.88 & 2008 Dec 04 \\
A2029 & 4977 &  ACIS-S &0.078 & 77.9 & 57.0 & 2004 Jan 08 \\
A2033 & 15167 & ACIS-I & 0.078 & 8.97 & 8.59 & 2013 May 22 \\
A2052 & 5807 & ACIS-S & 0.036 & 127.0 & 101.0 & 2006 Mar 24 \\
 & 10478 & ACIS-S & 0.036 &  119.1 & 101.5 & 2009 May 25 \\
A2063 & 4187 &ACIS-I & 0.034 & 8.8 & 1.4 & 2003 Apr 20 \\
A2110 & 15160 &ACIS-I &  0.098 & 7.98 & 7.00 & 2013 Aug 29 \\
A2199 & 10748 & ACIS-I & 0.031 & 40.6 & 36.2 & 2009 Nov 19 \\
A2204 & 7940 & ACIS-I & 0.151 & 77.1 & 63.7 & 2007 Jun 06 \\
A2355 & 15097 &ACIS-I & 0.231 & 19.8 & 14.7 & 2013 Nov 05 \\
A2390 & 4193 & ACIS-S & 0.233 & 95.1 & 69.0 & 2003 Sep 11 \\
A2415 & 12272 & ACIS-I & 0.057 & 9.92 & 7.90 & 2010 Sep 24 \\
A2597 & 7329 &  ACIS-S & 0.083 & 60.1 & 43.6 & 2006 May 04 \\
A262 & 7921 &ACIS-S &  0.017 & 111.0 & 92.1 & 2006 Nov 20  \\
 & 2215 & ACIS-S & 0.017 & 28.7 & 24.4 & 2001 Aug 03 \\
A2626 & 16136 & ACIS-S & 0.055 & 111.0 & 86.3 & 2013 Oct 20 \\
A2634 & 4816 & ACIS-S & 0.030 & 49.5 & 39.7 & 2004 Aug 31 \\
A2665 & 12280 & ACIS-I & 0.057 & 9.92 & 5.92 & 2011 Jan 17 \\
A2667 & 2214 & ACIS-S & 0.235 & 9.65 & 9.05 & 2001 Jun 19 \\
A3017 & 15110 & ACIS-I & 0.220 & 14.9 & 12.0 & 2013 May 01 \\
A3526 & 16223  & ACIS-S & 0.010 & 179.0 & 145.0 & 2014 May 26 \\
A3528S & 8268 & ACIS-I & 0.057 & 8.08 & 8.07 & 2007 Mar 20 \\
A3581 & 12884 & ACIS-S & 0.022 & 84.5 & 68.7 & 2011 Jan 03  \\
 & 1650 & ACIS-S & 0.022 & 7.2 & 7.0 & 2001 Jun 07 \\
A3695 & 12274 & ACIS-I &0.089 & 9.87 & 9.37 & 2010 Aug 17  \\
A4059 & 5785 & ACIS-S &0.049 & 92.1 & 83.3 & 2005 Jan 26  \\
 & 897 &ACIS-S & 0.049 & 40.7 & 6.5 & 2000 Sep 24  \\
A478 & 1669 & ACIS-S & 0.086 & 42.4 & 31.2 & 2001 Jan 27 \\
A496 & 4976 &ACIS-S &  0.033 & 75.1 & 36.5 & 2004 Jul 22 \\
AS1101 & 11758 & ACIS-I & 0.056 & 97.7 & 79.6 & 2009 Aug 24 \\
AS780 & 9428 & ACIS-S & 0.234 & 39.6 & 34.6 & 2008 Jun 16 \\
AS851 & 11753 & ACIS-I & 0.010 & 72.6 & 64.7 & 2009 Aug 19 \\
Hercules & 5796 & ACIS-S & 0.155 & 47.5 & 43.2 & 2005 May 09 \\
Hydra & 4970 & ACIS-S & 0.055 & 98.8 & 86.7 & 2004 Oct 22 \\
RXJ0058.9+2657 & 6830 & ACIS-I & 0.048 & 94.4 & 78.2 & 2006 Sep 02 \\
RXJ0107.4+3227 & 2147 & ACIS-S & 0.018 & 44.4 & 33.2 & 2000 Nov 06 \\
RXJ0123.6+3315& 2882 &ACIS-I &  0.017 & 43.6 & 33.9 & 2002 Jan 08 \\
RXJ0341.3+1524 & 4182 &ACIS-I & 0.029 & 23.5 & 22.9 & 2003 Mar 11 \\
RXJ0352.9+1941 & 10466 &ACIS-S & 0.109 & 27.2 & 24.3 & 2008 Dec 18 \\
RXJ0439.0+0520 & 9369 & ACIS-I & 0.245 & 19.9 & 18.7 & 2007 Nov 12 \\
RXJ0751.3+5012 & 15170 & ACIS-I& 0.024 & 97.7 & 74.7 & 2013 May 14 \\
RXJ0819.6+6336 & 2199 & ACIS-S & 0.119 & 14.9 & 11.3 & 2000 Oct 19 \\
RXJ1050.4-1250 & 3243 &ACIS-S &  0.015 & 29.5 & 22.6 & 2002 Nov 05 \\
RXJ1304.3-3031 & 4998 & ACIS-I & 0.010 & 15.0 & 13.6 & 2004 Feb 15 \\
RXJ1315.4-1623 & 9399 & ACIS-S &0.009 & 82.7 & 66.9 & 2008 Mar 07 \\
 & 17196 &ACIS-S & 0.009 & 88.9 & 80.9 & 2015 May 11 \\ 
RXJ1320.1+3308  & 6941 & ACIS-S & 0.038 & 38.6 & 31.2 & 2005 Nov 01 \\
RXJ1501.1+0141 & 12952 &ACIS-S & 0.007 & 143.0 & 126.0 & 2011 Apr 05 \\
 & 9517 & ACIS-S & 0.007 & 98.8 & 82.0 & 2008 Jun 05 \\
RXJ1504.1-0248 & 5793 & ACIS-I & 0.217 & 39.2 & 30.7 & 2005 Mar 20 \\
RXJ1506.4+0136 & 7923 & ACIS-I &  0.006 & 90.0 & 76.4 & 2007 Jun 12 \\
RXJ1522.0+0741 & 900 & ACIS-I & 0.045 & 57.3 & 53.2 & 2000 Apr 03 \\
RXJ1524.2-3154 & 9401 & ACIS-S & 0.102 & 40.9 & 29.4 & 2008 Jan 07 \\
RXJ1539.5-8335 & 8266 &  ACIS-I & 0.076 & 7.99 & 7.64 & 2007 Jun 24 \\
RXJ1558.4-1410 & 9402 &ACIS-S &  0.097 & 40.1 & 31.4 & 2008 Apr 09 \\
RXJ1604.9+2356 & 9423 &ACIS-S & 0.032 & 74.5 & 56.9 & 2008 May 18\\
RXJ1715.3+5725 & 4194 & ACIS-I & 0.028 & 47.3 & 25.9 & 2003 Sep 17 \\
RXJ1720.1+2638 & 4361 & ACIS-I & 0.161 & 25.7 & 15.6 & 2002 Aug 19 \\
RXJ1750.2+3504 & 12252 & ACIS-I &  0.171 & 19.8 & 16.8 & 2010 Oct 15 \\
RXJ1844.1+4533 & 5295 & ACIS-I & 0.092 & 30.7 & 25.1 & 2004 Jan 29 \\
RXJ2129.6+0005 & 9370 &ACIS-I & 0.235 & 29.5 & 27.0 & 2009 Apr 03 \\
Z1665 & 15161 & ACIS-I  & 0.031 & 9.95 & 7.11 & 2013 Feb 27 \\
Z235 & 11735 & ACIS-S & 0.083 &  19.8  & 13.3 & 2009 Sep 06 \\
Z3146 & 9371 & ACIS-I & 0.291 &  40.2  & 28.0 & 2008 Jan 18 \\
Z7160 & 4192 & ACIS-I & 0.258 & 91.9 & 79.3 & 2003 Sep 05 \\
Z808 & 12253 & ACIS-I & 0.169 & 18.8 & 14.3 & 2010 Oct 06 \\
Z8193 & 14988 & ACIS-S & 0.175 & 18.2 & 11.8 & 2013 Oct 07 \\
Z8276 & 11708 & ACIS-S & 0.075 & 45.4 & 38.9 & 2009 Nov 26 
\label{tab:table1}
\enddata
\end{deluxetable*} 

\clearpage

\section{Sample} \label{sec:obs}
\vspace*{2mm}

\begin{figure}
\centering
\hspace*{-0.5cm}
\includegraphics[width=0.52\textwidth]{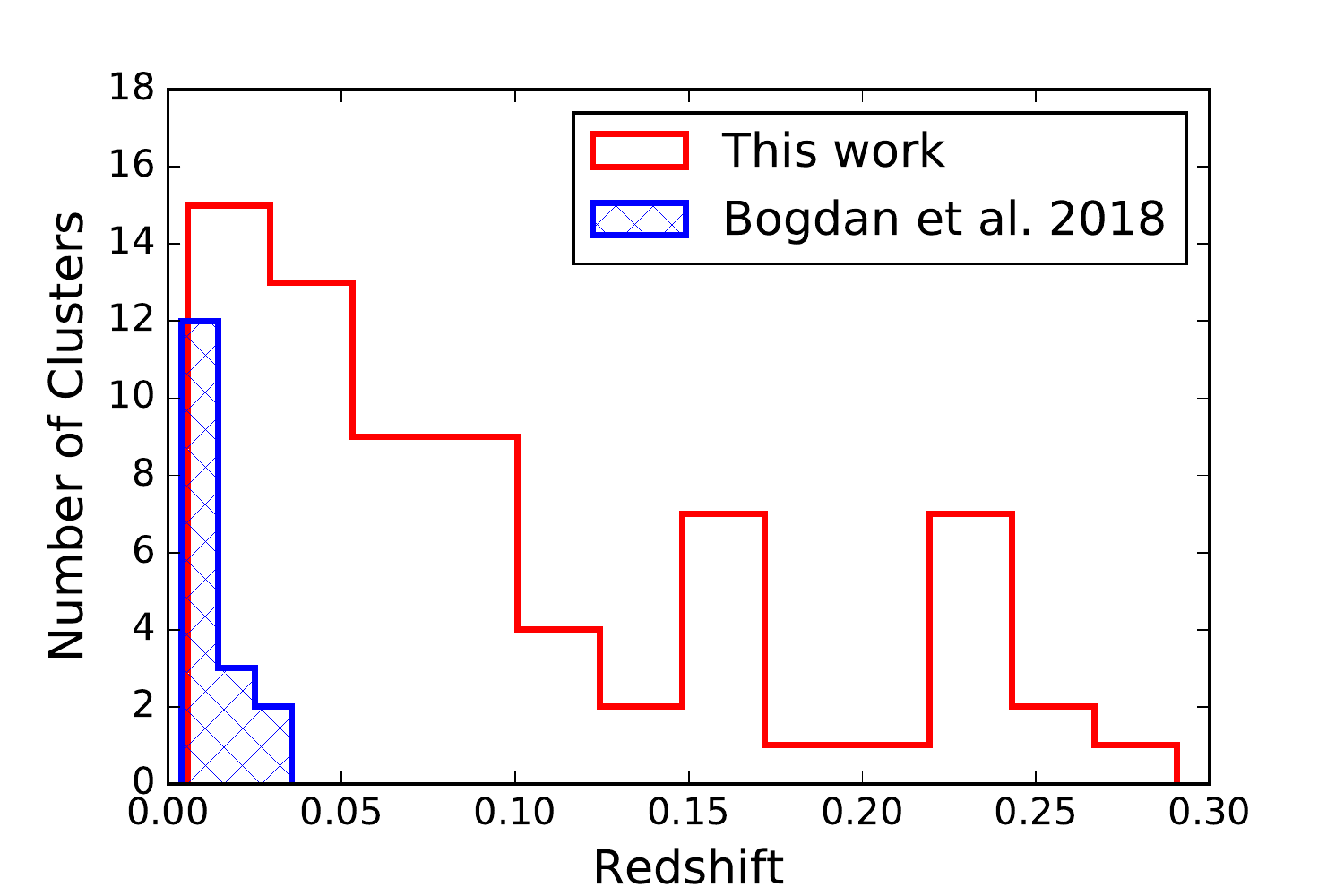} 
\caption{Redshift distribution for two cluster samples. The hatched (blue) distribution is the clusters in the sample studied by \citet{bogdan17}. The empty (red) distribution shows the clusters studied in this work. Note that the galaxy clusters investigated in this work have significantly higher redshifts than those in \citet{bogdan17}.} \label{fig:reddist}
\end{figure}

To study a larger set of galaxy clusters with indirect BH mass measurements, we rely on \cite{mezcua17} who investigated the Fundamental Plane relationship for BCGs. \cite{mezcua17} drew their sample from \cite{hogan15}, who examined the radio properties of X-ray selected BCGs from a parent sample of three \textit{ROSAT} X-ray catalogs. This sample includes galaxy clusters with redshifts in the range of $0.006 < z < 0.29$. By utilizing \textit{Chandra} X-ray and \textit{VLBI} radio observations, \citet{mezcua17} probed the Fundamental Plane relationship. They inferred the BH masses of the galaxies using the K-band luminosities of the BCGs. They concluded that the BHs were overly massive compared to the galaxy\textquotesingle s stellar mass. However, they did not expand their analysis to include the effects of the cluster. In this work, we utilize the Fundamental Plane relationship from \citet{merloni03}, together with the X-ray and radio luminosities from \citet{mezcua17}, in order to calcualte the BH masses.

In Figure \ref{fig:reddist} we show the redshift distribution for our sample as well as that for the sample studied by \citet{bogdan17}. Given the larger redshift range of the present sample, we can explore a larger volume, hence more massive clusters can be studied. However, while dynamically measured values for BH mass are preferable in constraining accurate relationships, they are limited to nearby and massive BHs. Therefore, without the use of tracers we would not be able to significantly increase the sample size and explore higher redshift, and, hence more massive clusters. 

\section{Data Analysis} \label{sec:analysis}
\label{sec:analysis}
\subsection{\textit{Chandra} Data}
\subsubsection{Data Reduction}
\label{sec:reduction}
To study the ICM of the galaxy clusters, we utilized \textit{Chandra} X-ray observations. All the systems have been observed with \textit{Chandra}  -- as it was also used by \citet{mezcua17}. Although \textit{XMM-Newton} data is avaliable for some sources in our sample, in order to avoid calibration issues (e.g. see \citealt{schel15}) we opted to use only \textit{Chandra} data. The data analysis was performed using CIAO software version 4.9 and CALDB 4.7.6 \citep{frus06}. 

As the first step of the analysis, we reprocessed all \textit{Chandra} observations using the \textit{chandra\textunderscore repro} tool. Since we aim to study the diffuse emission, we must identify and remove bright point sources. To this end, we utilized the \textit{wavdetect} task, which correlates the data with a Mexican Hat wavelet function of different scales and generates a list of point sources.  The applied scales were the square root series of two from $\sqrt{2}$ to $8.0$. The point sources found by \textit{wavdetect} were then masked from the analysis of the diffuse X-ray emission. 

Once the point sources were excluded, we filtered the time periods that had high background due to flares. For each observation, we extracted light curves in the $2.3-7.3$~keV energy range and binned them with a time interval of 200 s. Light curves were extracted within this energy range, as \textit{Chandra} is most sensitive to flares in this band \citep{hick06}. We applied the \textit{deflare} tool with the parameter $nsigma=2$. Hence, we removed all time periods that were $\geq 2\sigma$ outliers from the mean, which resulted in exposures that were typically $15-25\%$ shorter than the original exposures (Table \ref{tab:table1}). Although we could be less conservative by applying a $3 \sigma$ clipping instead, we prefer to use a $2 \sigma$ clip. We use this clipping since it is more effective in excluding soft proton flares, and therefore more effective in avoiding bias in the spectral fit procedure.

To account for the sky and instrumental background components, we constructed the blank-sky background for each observation. We used the \textit{blanksky} tool to create the background event files. While the spectrum of the background remains invariable, its normalization exhibits variations. To account for this, we used the count rates in the $10-12$ keV energy range to re-scale the blank-sky background files. 

\startlongtable
\begin{deluxetable*}{cccccccc}
\centering
\tablecaption{Characteristics of the analyzed galaxy clusters, their BCGs, and BHs.}
\startdata
Name & $N_{\rm H}$ & $\log_{\rm 10}(M_{\rm BH}) $ &  $\log_{\rm 10}(M_{\rm bulge})$ & $kT$  & $M_{\rm 500}$ & $R_{\rm 500}$ & $R_{\rm frac}$ \\ 
& ($10^{20} \rm cm^{-2}$) &    ($M_{\odot}$)& ($M_{\odot}$) & (keV) &  ($10^{14}M_{\odot}$) & (Mpc) & \\
(1) & (2) & (3) & (4) & (5) & (6) & (7) & (8) \\ \hline 
A1204 & 1.38 & 9.01 \plm 0.62 & 11.77 \plm 0.02 &  2.51 \plm 0.31 & 1.67 \plm 0.18 & 0.716 \plm 0.016 & 1.00 \\
A1367\textsuperscript{*} & 2.39 &  $<11.19$ &  11.40 \plm 0.01  & 2.40 \plm 0.07 & 2.11 \plm 0.04 & 0.887 \plm 0.007 & 0.46 \\
A1446 & 1.5 & 10 \plm 0.08 & 11.83 \plm 0.02 &  3.17 \plm 0.25 & 2.49 \plm 0.17 & 0.867 \plm 0.012 & 0.68 \\
A1644\textsuperscript{*} & 5.19 &  $<10.78$ &  12.13 \plm 0.04  &2.90 \plm 0.11 & 2.14 \plm 0.07 & 0.0867 \plm 0.006 & 0.73  \\
A1664\textsuperscript{*} & 8.72 & $<10.58$ & 11.76 \plm 0.02 &  4.54 \plm 0.42 & 4.60 \plm 0.36 & 1.040 \plm 0.017 & 0.73 \\
A168\textsuperscript{*} &3.25 & $<9.14$ &  11.76 \plm 0.02 & 1.69 \plm 0.07 & 0.85 \plm 0.03 & 0.640 \plm 0.005 & 0.58 \\
A1763 &  0.92 & 9.94 \plm 0.11 & 12.37 \plm 0.05  &5.82 \plm 0.66 &  7.04 \plm 0.68 & 1.101 \plm 0.022  & 1.00 \\
A1795\textsuperscript{*} & 1.17 & $<10.83 $&  12.05 \plm 0.03 &6.07 \plm 0.35 & 7.56 \plm 0.37 & 1.302 \plm 0.014 & 0.69 \\
A1930 & 1.13 & 9.79 \plm 0.16 & 12.25 \plm 0.05 & 3.34 \plm 0.43 &  2.72 \plm 0.3 & 0.870 \plm 0.020 & 0.89 \\
A2009 &3.27 &  9.74 \plm 0.08 &  12.17 \plm 0.04 & 3.89 \plm 0.55 & 3.53 \plm 0.43 & 0.931 \plm 0.024 & 1.00 \\
A2029\textsuperscript{*} & 3.15 & $<9.38$ &  11.90 \plm 0.03 & 8.76 \plm 0.05 & 10.1 \plm 7.7 & 1.415  \plm 0.361 & 0.71\\
A2033 & 2.94 & 9.68 \plm 0.15 &  12.17 \plm 0.04 & 2.12 \plm 0.23 & 1.25 \plm 0.12 & 0.705 \plm 0.014  & 0.96 \\
A2052 & 2.85 & 11.23 \plm 0.12 &  11.93 \plm 0.03 &  3.01 \plm 0.05 & 2.57 \plm 0.04 & 0.936 \plm 0.005  & 0.68 \\
A2063\textsuperscript{*} & 3.04 & $<9.03$ &  11.76 \plm 0.02 & 3.39 \plm 0.50 & 2.90 \plm 0.44 & 0.974 \plm 0.049 & 0.61  \\
A2110& 2.39 & 8.88 \plm 0.16 &  11.91 \plm 0.03 & 2.63 \plm 0.38 & 1.52 \plm 0.21 & 0.738 \plm 0.021 & 1.00  \\
A2199\textsuperscript{*} & 0.89 & $<9.72 $&  11.86 \plm 0.02 &  3.25 \plm 0.09 & 2.59 \plm 0.06 & 0.940 \plm 0.005 & 0.38  \\
A2204\textsuperscript{*} &5.69 & $<9.77$ &  12.15 \plm 0.04 & 5.97 \plm 0.25 & 7.34 \plm 0.26 & 1.191 \plm 0.009 & 1.00 \\
A2355 &  4.75 & 10.29 \plm 0.26 &  12.21 \plm 0.05 & 6.17 \plm 1.03 & 7.77 \plm 1.11 & 1.135 \plm 0.034 & 1.00  \\
A2390\textsuperscript{*} & 6.89 & $<12.05$ &  12.33 \plm 0.06 & 11.62 \plm 0.97 & 22.9 \plm 1.64 & 1.626 \plm 0.024  & 0.89 \\
A2415 & 4.79 & 11.13 \plm 0.01 & 11.44 \plm 0.02  &  2.18 \plm 0.55 & 1.31 \plm 0.28 & 0.731 \plm 0.033 & 0.71  \\
A2597\textsuperscript{*} & 2.49 & $<10.65 $& 11.57 \plm 0.02 &  3.42 \plm 0.14 & 2.83 \plm 0.10 & 0.921 \plm 0.007 & 0.54  \\
A262 & 5.46 & 9.01 \plm 0.08 &  11.63 \plm 0.02 &  2.33 \plm 0.08 & 2.06 \plm 0.06 & 0.885 \plm 0.008 & 0.34  \\
A2626 & 4.33 & 9.71 \plm 0.01 &  11.98 \plm 0.03 & 3.20 \plm 0.14 & 2.53 \plm 0.10 & 0.911 \plm 0.007 & 0.75 \\
A2634 & 5.06 & 10.88 \plm 0.15 & 11.77 \plm 0.02 &  5.89 \plm 0.32 & 7.19 \plm 0.33 & 1.322 \plm 0.013 & 0.50 \\
A2665 & 6.04 & 8.58 \plm 0.15 &  11.99 \plm 0.03 &2.68 \plm 1.79 &  1.87 \plm 1.07 & 0.822 \plm 0.099 & 0.66 \\
A2667 & 1.65 & 10.35 \plm 0.27 &  11.97 \plm 0.04 & 7.05 \plm 1.09 & 9.77 \plm 1.29 & 1.222 \plm 0.034 & 1.00 \\
A3017 &2.09 &  9.76 \plm 0.17 &  11.73 \plm 0.02 & 4.71 \plm 0.84 & 4.90 \plm 0.75 & 0.983 \plm 0.031 & 1.00 \\
A3526\textsuperscript{*} & 8.1 &  $<10.55$ & 11.77 \plm 0.02 &   2.26 \plm 0.01 &1.40 \plm 0.01 & 0.781 \plm 0.001 & 0.82 \\
A3528S\textsuperscript{*} & 6.13 & $<8.77$& 12.11 \plm 0.04 &  2.06 \plm 0.24 & 1.19 \plm 0.12 & 0.707 \plm 0.015  &  0.88\\
A3581 & 4.25 & 10.41 \plm 0.05 &  11.48 \plm 0.01 & 1.62 \plm 0.05 & 1.63 \plm 0.03 & 0.814 \plm 0.005 & 0.65 \\
A3695 & 3.7 & 10.37 \plm 0.08 &  11.90 \plm 0.03 &  2.80 \plm 0.29 & 2.01 \plm 0.18 & 0.817 \plm 0.015 & 1.00 \\
A4059\textsuperscript{*} & 1.1 & $<9.38 $&  12.20 \plm 0.04 &  5.96 \plm 0.25 & 5.76  \plm 0.34 & 1.208 \plm 0.024 & 0.61 \\
A478\textsuperscript{*} & 26.8 &  $<9.90 $ &  11.93 \plm 0.03 & 7.82 \plm 0.26 & 8.48 \plm 0.93 & 1.327 \plm 0.028 & 0.71 \\
A496\textsuperscript{*} &  4.8 & $<10.77$ &  12.00 \plm 0.03 & 7.51 \plm 0.43 & 10.9 \plm 0.53 & 1.513 \plm 0.016  & 0.58\\
AS1101\textsuperscript{*} &  1.83 & $<9.42$ & 11.92 \plm 0.03 & 1.37 \plm 0.01 &  0.59 \plm 0.01 & 0.560 \plm 0.001 & 0.85  \\
AS780 & 7.72 & 11.14 \plm 0.01 & 12.25 \plm 0.05 &  5.57 \plm 0.68 & 6.52 \plm 0.68 & 1.068 \plm 0.023  & 1.00 \\
AS851 & 4.96 &  9.76 \plm 0.08 & 11.61 \plm 0.02 &  0.73 \plm 0.01 & 0.20 \plm 0.01 & 0.412 \plm 0.001 & 0.77 \\
Hercules\textsuperscript{*} & 6.33 & $<10.15$ & 12.04 \plm 0.03 & 3.25 \plm 0.23 &  2.60 \plm 0.16 & 0.840 \plm 0.010 & 1.00  \\
Hydra & 4.84 & 10.99 \plm 0.62 &  11.78 \plm 0.02 &  4.13 \plm 0.09 & 3.92 \plm 0.07 & 1.054 \plm 0.004 & 0.69  \\
RXJ0058.9+2657 & 5.73 &  10.69 \plm 0.08 &  11.91 \plm 0.03 & 1.33 \plm 0.07 & 0.57 \plm 0.03 & 0.556 \plm 0.006 & 0.79  \\
RXJ0107.4+3227 & 5.41 &9.88 \plm 0.08 &  11.65 \plm 0.02 & 6.07 \plm 0.76 & 7.56 \plm 0.81 & 1.361 \plm 0.031 & 0.33  \\
RXJ0123.6+3315 & 5.23 &  8.28 \plm 0.01 &  11.69 \plm 0.02 & 1.32 \plm 0.03 & 0.56 \plm 0.01 & 0.570 \plm 0.002 & 0.26   \\
RXJ0341.3+1524 & 16.17 & 9.53 \plm 0.08 &  11.19 \plm 0.02 & 1.72 \plm 0.02 & 0.87 \plm 0.01 & 0.656 \plm 0.002 & 0.45 \\
RXJ0352.9+1941 & 26.83 & 8.97 \plm 0.06 & 11.85 \plm 0.03 &   2.19 \plm 0.15 & 1.96 \plm 0.86 & 0.798 \plm 0.019 & 0.75  \\
RXJ0439.0+0520 & 10.3 &  11.95 \plm 0.12 & 12.10 \plm 0.04 &  3.19 \plm 0.43 & 2.51 \plm 0.29 & 0.770 \plm 0.019  & 1.00 \\
RXJ0751.3+5012\textsuperscript{*} & 5.09 &  $<9.21$ &  11.33 \plm 0.01 & 0.99 \plm 0.07 & 0.34 \plm 0.02 & 0.482 \plm 0.006 & 0.40  \\
RXJ0819.6+6336 & 4.16 &  8.47 \plm 0.12 &  12.12 \plm 0.04 & 3.55 \plm 0.68 & 3.01 \plm 0.50 & 0.911 \plm 0.032 & 0.75  \\
RXJ1050.4-1250 &4.5 &  7.95 \plm 0.09 &  11.39 \plm 0.01 & 0.96 \plm 0.06 & 0.33 \plm 0.02 & 0.478 \plm 0.005 & 0.53 \\
RXJ1304.3-3031\textsuperscript{*} &6.01 &  $<9.32$ &  11.58 \plm 0.02 & 0.95 \plm 0.03 & 0.32 \plm 0.01 & 0.477 \plm 0.003 & 0.91 \\
RXJ1315.4-1623 & 4.94 & 9.26 \plm 0.08 &  11.44 \plm 0.01 & 1.20 \plm 0.03 & 1.43 \plm 0.01 & 0.789 \plm 0.003 & 0.30  \\
RXJ1320.1+3308 &1.05 &  9.32 \plm 0.12 &  11.46 \plm 0.01 & 1.04 \plm 0.04 & 0.37 \plm 0.01 & 0.487 \plm 0.003 & 0.71 \\
RXJ1501.1+0141 & 4.25 & 8.06 \plm 0.08 &  11.21 \plm 0.02 & 0.59 \plm 0.09 & 1.22 \plm 0.02 & 0.750 \plm 0.004 & 0.23  \\
RXJ1504.1-0248\textsuperscript{*}&  6.1  & $<10.69$ &  11.93 \plm 0.03 & 6.99 \plm 0.57 & 9.63 \plm 0.67 & 1.233 \plm 0.018 & 1.00  \\
RXJ1506.4+0136 & 4.24 & 8.97 \plm 0.23 &  11.31 \plm 0.01 &  0.89 \plm 0.01 & 0.28 \plm 0.01 & 0.460 \plm 0.001 & 0.47  \\
RXJ1522.0+0741\textsuperscript{*} & 3.05 &  $<8.91$ & 11.63 \plm 0.02 &  2.96 \plm 0.10 &  2.22 \plm 0.06 & 0.880 \plm 0.005 & 0.65  \\
RXJ1524.2-3154\textsuperscript{*} &8.44 & $<10.55$ & 11.87 \plm 0.03 & 3.00 \plm 0.21 &  2.26 \plm 0.13 & 0.841 \plm 0.010 & 0.70 \\
RXJ1539.5-8335 & 7.68 & 10.19 \plm 0.15 &  12.10 \plm 0.04 & 1.76 \plm 0.28 & 0.91 \plm 0.1 & 0.635 \plm 0.018 & 0.67  \\
RXJ1558.4-1410 & 11.47 &  12.23 \plm 0.15 & 12.19 \plm 0.04 &  3.84 \plm 0.22 & 3.45 \plm 0.17 & 0.972 \plm 0.010 & 0.64  \\
RXJ1604.9+2356 & 4.99 & 10.77 \plm 0.06 &  11.85 \plm 0.02 & 2.27 \plm 0.28 & 1.41 \plm 0.15 & 0.766 \plm 0.017 & 0.53 \\
RXJ1715.3+5725 & 2.6 & 9.96 \plm 0.55 &  11.75 \plm 0.02 & 1.37 \plm 0.02 & 0.59 \plm 0.01 & 0.575 \plm 0.001 & 0.39 \\
RXJ1720.1+2638\textsuperscript{*} & 3.89 & $<9.15$ &  12.13 \plm 0.04 & 5.39 \plm 0.46 &  6.17 \plm 0.45 & 1.115 \plm 0.017 & 1.00 \\
RXJ1750.2+3504 & 3.12 & 11.08 \plm 0.15 &  12.17 \plm 0.04 & 3.00 \plm 0.49 & 2.27 \plm 0.31 & 0.791 \plm 0.023 & 1.00 \\ 
RXJ1844.1+4533 & 6.32 & 10.72 \plm 0.01 & 11.97 \plm 0.03 & 1.72 \plm 0.11 &  0.88 \plm 0.01 & 0.618 \plm 0.007 & 1.00 \\
RXJ2129.6+0005\textsuperscript{*} &4.16 &  $<9.76$ & 12.16 \plm 0.04 & 4.30 \plm 0.36 &   4.19 \plm 0.3 & 0.921 \plm 0.014 & 1.00  \\
Z1665\textsuperscript{*} &2.74 &  $<9.09$&  11.65 \plm 0.02 &  1.48 \plm 0.26 & 0.68 \plm 0.1 & 0.601 \plm 0.019 & 0.56  \\
Z235\textsuperscript{*} & 3.91 & $<10.65$ & 11.95 \plm 0.03 &  3.45 \plm 0.35 & 2.87 \plm 0.25 & 0.926 \plm 0.017 & 0.55 \\
Z3146\textsuperscript{*} &2.93 & $<8.73$ &  12.19 \plm 0.06 & 5.59 \plm 0.55 & 6.57 \plm 0.55 & 1.024 \plm 0.018  & 1.00 \\
Z7160\textsuperscript{*} &3.22 & $<9.03$& 12.41 \plm 0.06  &  3.60 \plm 0.26 & 3.09 \plm 0.19 & 0.817 \plm 0.011 & 1.00 \\
Z808\textsuperscript{*} &7.55 & $<9.06$ &  11.95 \plm 0.03 &  3.03 \plm 0.58 & 2.30 \plm 0.38 & 0.797 \plm 0.027 & 1.00 \\
Z8193\textsuperscript{*} &2.31& $<11.33$&  12.28 \plm 0.05 & 4.47 \plm 0.71 & 4.47 \plm 0.61 & 0.989 \plm 0.028 & 0.72 \\
Z8276 & 3.66 & 10.53 \plm 0.54 & 11.89 \plm 0.03 & 3.79 \plm 0.23 &  3.38 \plm 0.17 & 0.985 \plm 0.011 & 0.94  
\label{tab:table2}
\enddata
\tablenotetext{*}{Sources where the BH mass is an upper limit}
\tablecomments{Columns are as follows: (1) Name of the galaxy cluster; (2) Line-of-sight column density to the cluster \citep{kal05}; (3) BH mass obtained from the Fundamental Plane relation \citep{merloni03}; (4) Stellar bulge mass of the BCG calculated using K-band luminosity and the mass-to-light ratio of 0.85 \citep{bell03}; (5) Best-fit temperature of the ICM; (6) $M_{\rm 500}$ mass inferred from the best-fit temperature in column (5) using the $kT-M_{\rm 500}$ relation of \citet{lov15}; (7) $R_{\rm 500}$ radius of the cluster; (8) The fraction of the $R_{\rm 500}$ radius included in the \textit{Chandra} field of view. The errors associated with the $M_{\rm 500}$ mass and $R_{\rm 500}$ radius were computed from the temperature uncertainties.}
\end{deluxetable*}

\subsubsection{Measuring the Temperature of the ICM}
\label{sec:temp}
To accurately measure the ICM temperature, we first identified the peak of the X-ray emission, which is considered to be the center of the cluster. To find the X-ray peak, we smoothed the $0.7-2$ keV band images with a Gaussian with a kernel size of 3. We searched for the maximum on this smoothed image, which defined the peak of the emission. We note that the center of the cluster may be slightly offset from the BCG \citep{hudson10}. Several clusters in our sample exhibit double peaked profiles, as these may be undergo mergers (see Section \ref{sec:discuss}). In these cases the second peak was masked from the observations.

After we identified the center of each cluster, we determined the $R_{\rm 500}$ radius. As the first step, we computed the signal-to-noise ratio in the $0.7-2$ keV band using concentric annuli. The first spectra were extracted within a radius where the signal-to-noise ratio peaked. After extracting the spectrum, we determined the initial temperature and an initial $M_{\rm 500}$\footnote{$M_{\rm 500}$ is the mass contained within the radius where the density of the cluster is $500$ times the critical density of the Universe} using the $kT - M_{\rm 500}$ relation by \cite{lov15}:
\begin{equation} \label{eqn:m500}
\log_{10}\Big(\frac{M_{\rm 500}}{5\times10^{13} \ h^{-1} \ M_{\odot}} \Big) = 1.71 \log_{10}\Big(\frac{kT}{\rm 2 \ keV}\Big) + 0.20
\end{equation}
Where $h$ is the reduced Hubble constant. This $M_{\rm 500}$ value was then used to calculate the $R_{\rm 500}$ radius.

Once the initial $R_{\rm 500}$ radius was retrieved, we followed the iterative process outlined in \cite{lov17} to measure the gas temperature of the ICM. Specifically, we extracted the source and background spectra from the $(0.15-0.75)R_{\rm 500}$ region, in order to exclude the central regions where there could be an extreme temperature gradient. These spectra were fit using \textit{XSpec} \citep{arn96}. From the fit we obtained a new temperature, and, hence $M_{\rm 500}$ value using the $kT - M_{\rm 500}$ relation in \citet{lov15}. This new $M_{\rm 500}$ and the inferred $R_{\rm 500}$ defined the new extraction annulus. We continued to iterate via this process until the temperature remained invariant within $5\%$. Although \citet{lov17} used the $M-Y_{\rm x}$ relation, in this work we rely on the $kT - M_{\rm 500}$ relation for consistency with \citet{bogdan17}. For most of the clusters, the extraction region lay within the field-of-view of the analyzed \textit{Chandra} observations. However, for a small sample of nearby clusters, only a relatively small fraction of the $R_{\rm 500}$ were included in the field-of-view. For these systems, we utilized multiple observations to increase the coverage of the clusters. In Appendix A we show all the \textit{Chandra} images of the clusters in the $0.7 - 2.0$ keV band as well as the annulus which the final spectra was extracted from defined by the $R_{\rm 500}$ value.

Spectra along with the corresponding response files for the source and background data were extracted using the \textit{specextract} and \textit{dmextract} tools, respectively. The background spectra were renormalized using the count rate ratios observed in the $10 -12$ keV band, and the re-normalized background spectrum was subtracted from the source spectrum. The background-subtracted source spectrum was grouped by count number with each bin requiring a minimum of $15$ counts. This final spectrum was used to fit a model and determine the ICM temperature of the galaxy cluster. 

We used the spectral analysis software, \textit{XSpec}, to fit a model to the data \citep{arn96}. Fitting was performed in the $0.7-5 $~keV energy band. We fit the emission with an \textit{apec} model that describes collisionally ionized thermal plasma. We allowed the abundance, temperature and normalization to vary. For the abundances we used the table of \citet{and89}  In addition, we included photoelectric absorption, whose value were the weighted average from the Leiden/Argentine/Bonn (LAB) survey \citep{kal05}.  This procedure resulted in acceptable fits for most clusters in our sample. However, for two clusters, A478 and J0352.9+1941, the column densities provided by the LAB survey were underestimated. We obtained significant residuals at $<1$ keV for the spectra of these clusters. The LAB survey underestimates column densities, and this underestimation becomes strongly signficant for high values  (excess of $10^{21}$), such as for these clusters. Therefore, for these two systems we allowed the column density to vary during the fitting process. Allowing this parameter to vary results in an acceptable fit.

In Table \ref{tab:table2} we list the obtained $R_{\rm 500}$ and $M_{\rm 500}$ values as well as the best-fit ICM temperatures.

\begin{figure*}
\centering
\hspace*{-1.cm}
\includegraphics[width=0.70\linewidth]{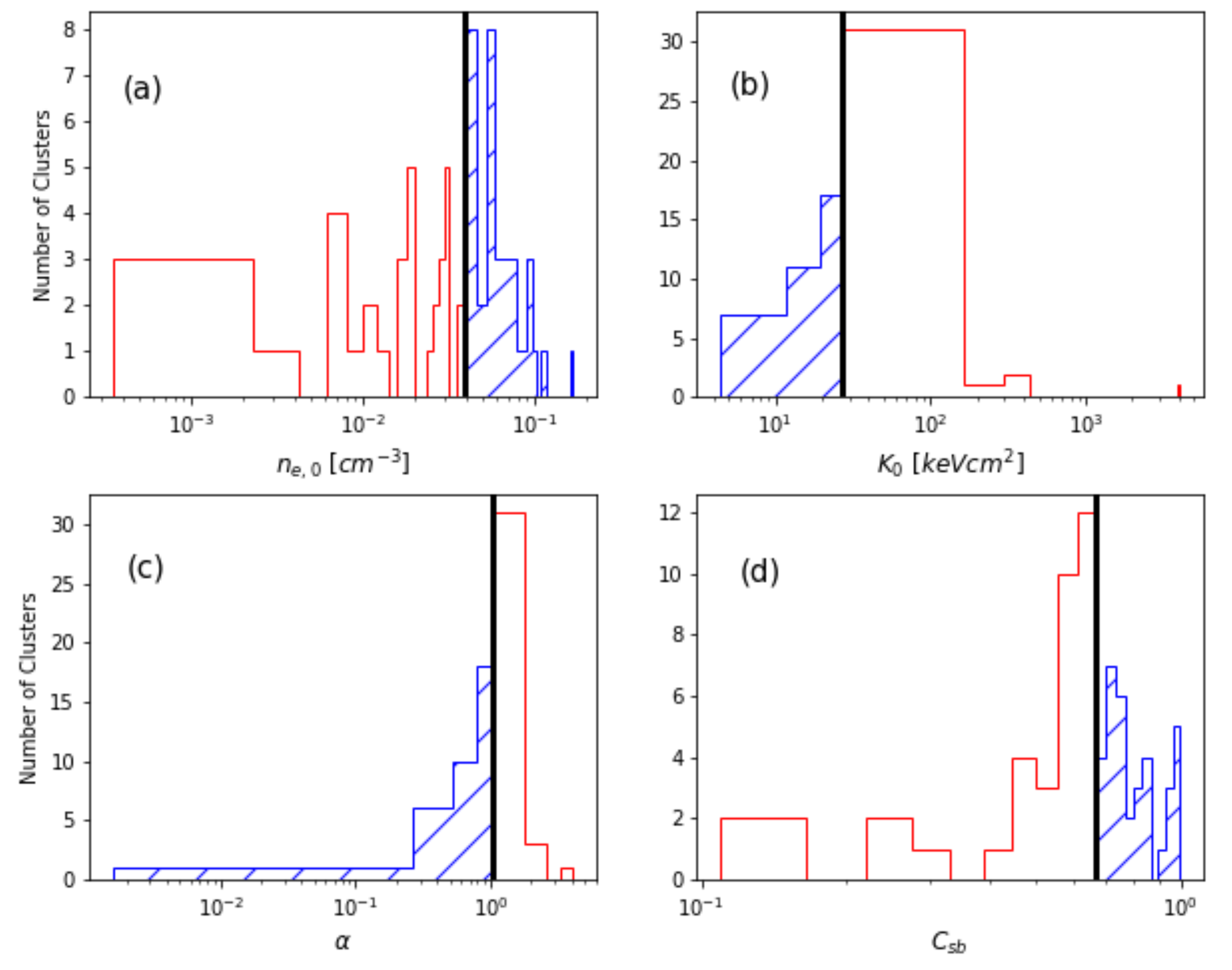}
\caption{The distributions of relaxed and disturbed clusters based on the four parameters from \citet{hudson10}; (a) central electron density, \neo, (b) central entropy, \Ko, (c) cuspiness, $\alpha$, and (d) central surface brightness, \csb. The relaxed clusters, which have similar properties to CCs are shown in blue hatch.  The disturbed clusters, which can be likened to NCC clusters, are shown in red. The black lines indicate the median for each parameter where the clusters split into the two subsamples.} \label{fig:histcc}
\end{figure*}

\subsection{Cool Core and Non-Cool Core Clusters}
\label{sec:ccncc}

Given our larger sample size, we aim to explore whether the \MbhT relationship exhibits variations for different types of clusters. Specifically, we split our sample into cool core (CC) and non-cool core (NCC) clusters. Due to their dynamics, CC clusters generally have greater gas inflow to the center \citep{hudson10}. This process may increase the growth rate of the BH residing in the BCG. This, in turn, could lead to differences in the \MbhT relation for CC and NCC clusters. 

There are several parameters that can be used to distinguish between these two types of cluster and the debate is ongoing as to which parameter is the most accurate. Therefore, we used four parameters to split the galaxy cluster samples into CC and NCC clusters to examine whether there is a difference in the \MbhT relationship. These parameters are central electron density, central entropy, cuspiness, and surface brightness density \citep{hudson10}. Instead of using a value from literature, the median value of each of the parameters was used to split our sample. We split our clusters in this way because, if we were to use a value from the literature, then we would need our data to be calibrated perfectly with the original work. As this is difficult, by using the median value instead of classifying the clusters definitively as a CC (NCC), we simply make a distinction between more relaxed (disturbed) systems. Therefore from now on we will refer to the two populations as more relaxed, e.g.  clusters showing CC properties, and more disturbed (clusters showing NCC properties).  

To calculate these parameters, we constructed surface brightness profiles in the $0.7-2$ keV band for each cluster. For each cluster, we created 38 concentric annuli with a minimum of $10 $ pixels and a maximum of $200 $ pixels. To account for the background, we used the blank-sky images in the same energy range and with the same annuli. The obtained background subtracted surface brightness profiles were fit with a double-beta model \citep{hudson10}: 
\begin{equation} \label{eqn:doublebeta}
\Sigma = A_{0,1}\Big[1 + \Big(\frac{r}{r_{c,1}}\Big)^2\Big]^{-3\beta_1 + 0.5} + A_{0,2}\Big[1+\Big(\frac{r}{r_{c,2}}\Big)^2\Big]^{-3\beta_2 + 0.5}
\end{equation}
where $A_{0,1}$ and $A_{0,2}$ are the amplitudes, $r_{c,1}$ and $r_{c,2}$ are the core radii. The values for amplitude, core radii and $\beta_{1,2}$ were fit for each cluster. Based on these, we computed the four parameters to define the relaxed and disturbed cluster sub-samples. The obtained distribution of these sub-samples are shown in Figure \ref{fig:histcc}.

\subsubsection{Central Electron Density}

To calculate the central electron density (\neo) from the double-beta model, we use: 
\begin{equation} \label{eqn:centralneo}
n_{e,0} = \Big[n_{e,1}^2\Big(1+\Big(\frac{r}{r_{c,1}}\Big)^2 \Big)^{-3\beta_1} + n_{e,2}^2\Big(1+\Big(\frac{r}{r_{c,2}}\Big)^2\Big)^{-3\beta_2}\Big]^{\frac{1}{2}} 
\end{equation}
where $n_{e,1}$ and $n_{e,2}$ are the central electron densities for each component in the double-beta model. The calculation of these quantities depends upon $n_0$, i.e. the central density:
\begin{equation}  \label{eqn:centraldense}
n_{0} = \Big(\frac{10^{14}4\pi D_AD_L\zeta N}{EI}\Big)^{0.5} 
\end{equation}
where $N$ is the normalization obtained from the best-fit \textit{apec} model. To calculate the value of $n_0$, we used $0.048R_{500}$ as the central extraction region \citep{hudson10}. The electron densities $n_{e,1}$ and $n_{e,2}$ for the double-beta model were calculated following \citet{hudson10}.

The value of \neo could be used to classify clusters as either disturbed or relaxed. The median which we took as a``splitting" value was \neo $= 0.040 \ \rm{cm^{-3}}$. Clusters with a density less than this value were classified as disturbed systems, as a hotter central temperature indicates a lower central density which is characteristic of a NCC system.

\subsubsection{Central Entropy}

To split the clusters based on their central entropy (\Ko), we compute this parameter using: 
\begin{equation} \label{eqn:k0}
K_0 = k_{B}T_0n_{e,0}^{-\frac{2}{3}}
\end{equation}
where $T_0$ is the central temperature, which is computed for the $0.048R_{500}$ region, thereby maintaining consistency with the normalization in Equation \ref{eqn:centraldense}. The  ``splitting" value for this parameter is \Ko $ =  26.59 \ \rm{keV \ cm^2}$. If  \Ko was less than this value then the cluster would be classified as a relaxed cluster as the gas at the center is cooler and hence less perturbed.


\begin{figure*}
\centering
\includegraphics[width=0.96\textwidth]{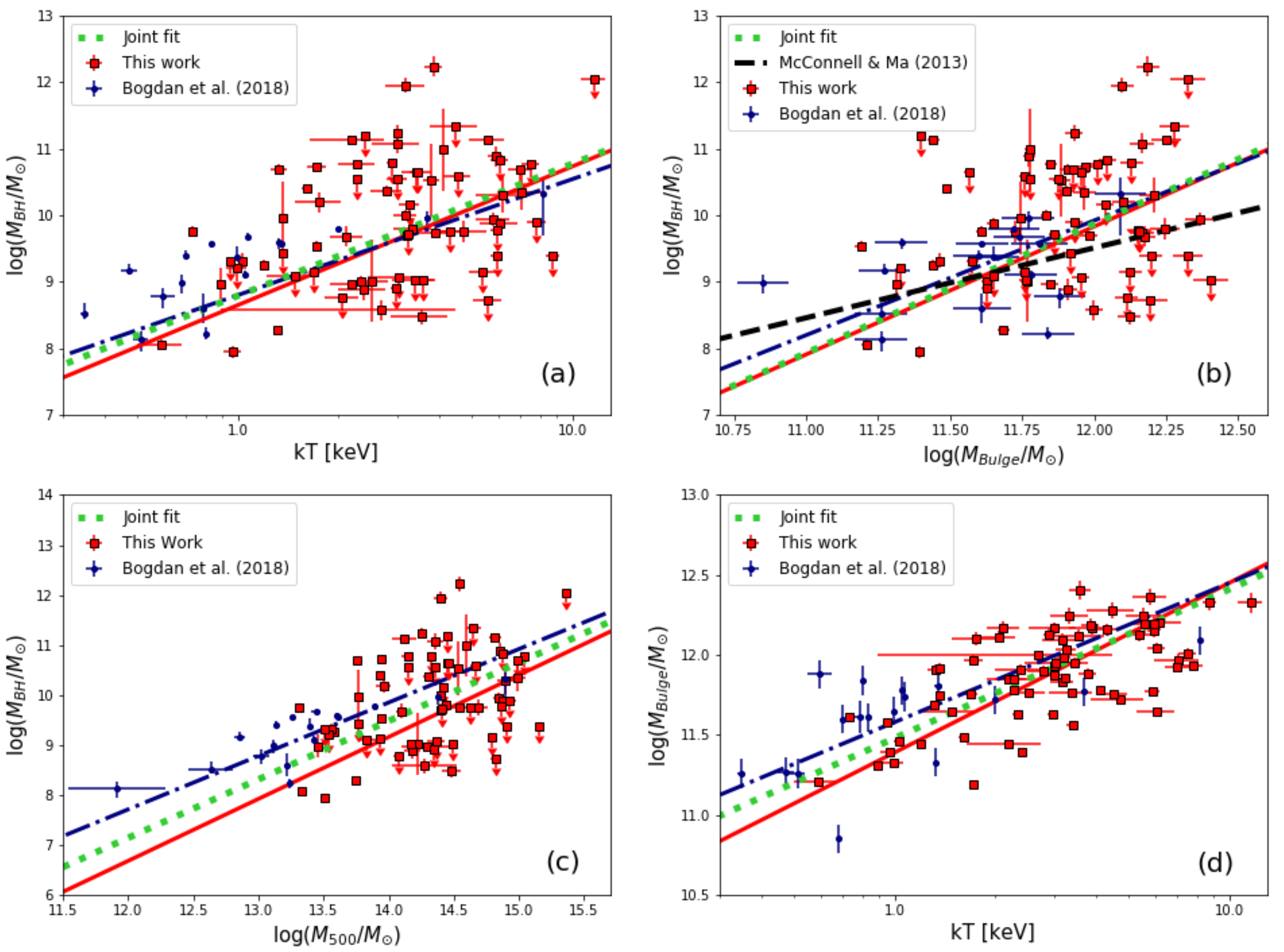}
\caption{Scaling relations investigated in this paper: (a) BH mass -- cluster temperature relation, (b) BH mass -- bulge mass relation, (c) BH mass -- cluster mass relation, and (d) bulge mass -- cluster temperature relation. The red squares are from this work with downward arrows indicating upper limits in BH mass. The blue circles are for the BGGs/BCGs studied in \citet{bogdan17}. The best-fitting lines for the data is shown with the solid line and were all calculated with BCES\textunderscore REGRESS code. For the fitting procedure the upper limits were included. The dot-dashed line in the relation from \citet{bogdan17}, the dotted line is from a joint fit based on the sample of this work and \citet{bogdan17} data, and in panel (b) the dashed line is the \MbhM relation from \citet{mccon13}.} \label{fig:scatt}
\end{figure*}
\begin{table*}
\caption{Best-fit parameters on linear regression for our data} \label{tab:params}
\centering
\begin{tabular}{ccccccc}
Relation & $\alpha$ & $\beta$ & $\sigma_{intrin}^x$ & $\sigma_{intrin}^y$ & r\textsuperscript{*} & $ \rho\textsuperscript{**} $ \\ \hline
\MbhT & -0.35 \plm 0.21 & 2.08 \plm 0.43 & 0.35 & 0.91 & 0.26 & 0.24 \\
\MbhM & -1.09  \plm 0.39 & 1.92 \plm 0.45 & 0.49 & 0.96 & 0.26 & 0.24 \\
$M_{\rm BH} - M_{\rm 500}$ & -1.08 \plm 0.25 & 1.24 \plm 0.19 & 0.69 & 0.99 & 0.18 & 0.20 \\
$M_{\rm Bulge}-$kT & 0.39 \plm 0.05 & 1.06 \plm 0.08 & 0.23 & 0.22 & 0.66 & 0.62 \\
\hline
\multicolumn{7}{l}{\textsuperscript{*}\footnotesize{Pearson Correlation Coefficient}}\\
\multicolumn{7}{l}{\textsuperscript{**}\footnotesize{Spearman Correlation Coefficient}}
\end{tabular}
\end{table*}

\subsubsection{Cuspiness}
\label{sec:cuspiness}

The cuspiness parameter ($\alpha$) was suggested as a parameter for identifying CC clusters at large redshifts \citep{vikh07} and it is defined as:
\begin{equation} \label{eqn:cusp1}
\alpha = - \frac{d\log(n_e(r))}{d\log(r)}
\end{equation}
when $r=0.04R_{\rm 500}$. As our density function is based on the double-beta model we can recast Equation \ref{eqn:cusp1} as the following:
\begin{equation} \label{eqn:cusp2}
\alpha = 3r^2 \frac{\Sigma_{12}LI_2\beta_1r_{c,1}^{-2}b_1' + LI_1\beta_2r_{c,2}^{-2}b_2'}{\Sigma_{12}LI_2b_1 + LI_1b_2}
\end{equation}
where the core radii and the $\beta_{1,2}$ are the values found from fitting the double-beta model (Equation \ref{eqn:doublebeta}) to the surface brightness profile. The values $\rm LI_i$ are the line emission measure for model i and $\sigma_{12}$ is the ratio of the central surface brightness of model 1 to model 2. Finally, $b_i, b_i'$ are defined as:
\begin{equation} \label{eqn:b}
b_{i} = \Big(1 + \Big(\frac{r}{r_{c,i}}\Big)^2\Big)^{-3\beta_i}
\end{equation}
and
\begin{equation} \label{eqn:bprime}
b_i' = \Big(1+\Big(\frac{r}{r_{c,i}}\Big)^2\Big)^{-3\beta_i -1}
\end{equation}
where i=1,2. For further details we refer to \citet{hudson10}.

The median ``splitting" value for this parameter is $\alpha = 1.05$. If $\alpha > 1.05$ then the cluster is classified as disturbed. 

\subsubsection{Surface Brightness Density}

The final parameter we used to identify clusters as either CC or NCC was the surface brightness density, $C_{\rm sb}$. This parameter was first used by \citet{san08}, as a way to measure the excess of brightness at the core of a cluster. The \csb is defined as:
\begin{equation} \label{eqn:csb}
C_{sb} = \frac{\Sigma(r < 40 \rm kpc)}{\Sigma(r < 400 \rm kpc)}
\end{equation}
In other words, the \csb is the ratio of the integrated surface brightness within a radius $40$ kpc to that within a radius of $400$ kpc. For this parameter a median ``splitting'' value of \csb $=0.66$ is used where clusters with a \csb greater than this value are classified as relaxed clusters, as CC clusters are bright at the center \citep{fab84}.  

We note that the median values we use to split the samples into relaxed and disturbed clusters differ from the literature, but this is to be expected due to differences between our samples and the data used.

\section{Results} \label{sec:results}
\subsection{Correlations}
\label{sec:correlations}

In the top left and right panels of Figure \ref{fig:scatt}, we depict the observed relation between the  BH mass and the cluster temperature and stellar bulge mass, respectively. Note that the BH masses are inferred from the Fundamental Plane relation \citep{merloni03}. The errors in the BH masses were calculated using the uncertainties obtained for the X-ray and radio luminosities of the BHs. To infer the BH mass from the Fundamental Plane, it is necessary to detect the BHs both in radio and X-ray wavelengths. However, in the sample of \citet{mezcua17} several BHs remain undetected in the X-ray band. Therefore, for these BHs we compute upper limits on the BH mass. To derive the stellar mass of the BCGs, we rely on \citet{mezcua17}, who provides the K-band absolute magnitude of the galaxies. We convert the K-band luminosities to stellar mass using the K-band mass-to-light ratio of $0.85$, which is typical for massive elliptical galaxies \citep{bell03}. On the plots we add the data points from \citet{bogdan17}, which used nearby galaxy groups/clusters with dynamically measured BHs. In the bottom left and right panels of Figure \ref{fig:scatt}, we show the relations between the inferred BH mass against cluster mass inferred from the $kT - M_{\rm 500}$ relation \citep{lov15}, as well as the relation between bulge mass and best-fit cluster temperature, respectively. We note that for six clusters in our sample, the fraction of the $R_{\rm 500}$ radius covered by the field of view of \textit{Chandra} is $R_{\rm frac} < 0.4$. After examining these clusters, we concluded that the BHs associated with their BCGs are not preferentially over-massive or under-massive relative to other systems. As such, these data points do not introduce a bias in our results.  

\begin{table*}
\caption{Best-fit parameters on linear regression for joint fitting} \label{tab:paramsjoint}
\centering
\begin{tabular}{ccccccc}
Relation & $\alpha$ & $\beta$ & $\sigma_{intrin}^x$ & $\sigma_{intrin}^y$ & r\textsuperscript{*} & $ \rho\textsuperscript{**} $ \\ \hline
 \MbhT & -0.21 \plm 0.10 & 1.98 \plm 0.24 & 0.35 & 0.84 & 0.37 & 0.35 \\
 \MbhM & -1.07 \plm 0.26 & 1.94 \plm 0.32 & 0.53 & 0.96 & 0.31 & 0.28 \\
 $M_{\rm BH} - M_{\rm 500}$  & -0.82 \plm 0.12 & 1.16 \plm 0.10 & 0.72 & 0.92 & 0.31 & 0.31 \\
$M_{\rm Bulge}-$kT & 0.48 \plm 0.04 & 0.93 \plm 0.07 & 0.23 & 0.25 & 0.70 & 0.69 \\
\hline
\multicolumn{7}{l}{\textsuperscript{*}\footnotesize{Pearson Correlation Coefficient}}\\
\multicolumn{7}{l}{\textsuperscript{**}\footnotesize{Spearman Correlation Coefficient}}
\end{tabular}
\end{table*}

To compute the best-fit relations, we used the BCES\textunderscore REGRESS code \citep{ark96}. This linear regression code uses bivariate correlated errors and intrinsic scatter fitting method.  It is advantageous to use this over standard linear regression fitting as it allows for the errors in both the x and y measurements to be taken into account. We performed the fits in $\log-\log$ space and used the \textit{bisector} method. In order to perform a fitting that included the upper limited sources we adopted the following method. For the non detections (shown with arrows in Figure \ref{fig:scatt}), we calculated a lower limit on the BH mass and then assumed that the true BH mass could be represented by a random uniform distribution between the upper and lower limits. The lower limits were calculated using the \MbhM relation of \cite{mccon13}. We take the lower limits to be $3 \sigma$ below the predicted value from this relation. In order to implement the BCES\textunderscore REGRESS code, all variables need an error. Therefore the non detections, for the BH masses we assume that the errors are $25 \%$ of the range between the measured upper limit and the calculated lower limit. We tested different percentages for the errors but found no significant difference in the result. For the detected BH mass points, we assumed a Gaussian distribution between the error range centered on the measured value. With these two random distributions for the detections and non detections we implemented the fitting code via a Monte Carlo simulation and repeated $10^4$ times. The values quoted below and in Table \ref{tab:params} are the mean from all these realizations. We obtained the following relationships: 

\begin{equation} \label{eqn:mbht}
  \log_{10} \Big(\frac{M_{\rm BH}}{10^9M_{\rm \odot}}\Big) = -0.35 + 2.08 \log_{10}\Big(\frac{kT}{\rm 1 \ keV}\Big) 
\end{equation}
\begin{equation} \label{eqn:mbhm}
 \log_{10}\Big(\frac{M_{\rm BH}}{10^9M_{\rm \odot}}\Big) = -1.09 + 1.92 \log_{10}\Big(\frac{M_{\rm bulge}}{10^{11}M_{\odot}}\Big) 
\end{equation}
\begin{equation} \label{eqn:mbhcm}
 \log_{10} \Big(\frac{M_{\rm BH}}{10^9M_{\rm \odot}}\Big) = -1.08 + 1.45 \log_{10} \Big(\frac{M_{\rm 500}}{10^{13} M_{\rm \odot}}\Big) 
\end{equation}
\begin{equation} \label{eqn:mbulget}
 \log_{10} \Big(\frac{M_{\rm Bulge}}{10^{11}M_{\rm \odot}}\Big) = 0.39 + 1.06 \log_{10}\Big(\frac{kT}{\rm 1 \ keV}\Big) 
\end{equation}

The intrinsic scatter in both the x and y planes were calculated following \citet{lov15}: 
\begin{align}
\label{eqn:intrinscatter}
\begin{split}
\sigma_{\rm intrin}^y = \sqrt{(\sigma_{\rm tot}^y)^2 - (\sigma_{\rm stat}^y)^2 - (a^2(\sigma_{\rm stat}^x)^2)}
\\
\sigma_{\rm intrin}^x = \sqrt{(\sigma_{\rm tot}^x)^2 - (\sigma_{\rm stat}^x)^2 - (a^{-2}(\sigma_{\rm stat}^y)^2)}
\end{split}
\end{align}
Where $\sigma_{\rm tot}^{x,y}$ and $\sigma_{\rm stat}^{x,y}$ represent the total and statistical scatter along x and y, and the value, $a$, is the gradient calculated from the BCES\textunderscore REGRESS code. The best-fit relations, the scatter, and the Pearson and Spearman correlation coefficients are tabulated in Table 3.

The best-fit \MbhT relation is somewhat steeper and exhibits larger scatter than the relation found in \citet{bogdan17}.  We note that the bulge masses in this work and in that of \citet{bogdan17} were calculated using different methods. To make the \MbhM relations comparable, we re-computed the bulge masses for the sample of \citet{bogdan17} using the method applied in this work. The \MbhM relations are identical within $1\sigma$ uncertainties. The intrinsic scatter in the \MbhT and \MbhM relations are $\sigma_{\rm intrin}^y = 0.91$ and $0.96$ respectively, which exceeds those found in \citet{bogdan17} and is comparable with the scatter in the Fundamental Plane relation of $0.88$ found by \citet{merloni03}. 

In the bottom left panel of Figure \ref{fig:scatt} we plot the BH mass against the $M_{\rm 500}$ mass of each cluster. To convert the cluster temperature to $M_{\rm 500}$, we used the $kT - M_{\rm 500}$ relation of \citet{lov15} (Equation \ref{eqn:m500}). The \MbM relationship (Equation \ref{eqn:mbhcm}) is consistent to within $1\sigma$ with that found by \citet{bogdan17}. In the bottom right panel of Figure \ref{fig:scatt}, we show the relation between the stellar bulge mass and the cluster temperature. While this relation exhibits smaller scatter (see Table \ref{tab:params}), the best-fit relation and the correlation coefficients are comparable to that obtained in \citet{bogdan17}.

\begin{figure*}
\epsscale{1.1}
\includegraphics[width=\textwidth]{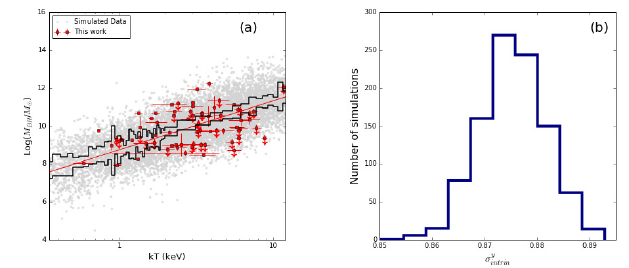}
\caption{Panel (a) shows the relationship between BH mass and cluster temperature. The red squares are the results from this work, downward arrows indicate upper limits. The light grey circles -- or cloud of data -- are the simulated data points from the Monte Carlo program. The red solid line is the relation from Equation \ref{eqn:mbht} calculated using BCES\textunderscore REGRESS code. The black lines either side the red one is the $1 \sigma$ contours from the line of best fit for the simulated data. Panel (b) shows the distribution in $\sigma_{\rm intrin}^y$ for the 1000 simulations.} \label{fig:simulation}
\end{figure*}

\subsection{Joint Fit Correlations}
\label{sec:joint}
The sample investigated in this work explores the high-mass end of the relations, while \citet{bogdan17} studied the low-mass end of the scaling relations. Therefore, here, we combine the two samples and perform joint fitting to establish relations, which are constrained across a broad range of galaxy groups and clusters with temperatures of $kT=0.4-12$ keV. We find the best-fitting equations to be: 
\begin{equation} \label{eqn:mbhtjoint}
 \log_{10} \Big(\frac{M_{\rm BH}}{10^9M_{\rm \odot}}\Big) = -0.21 + 1.98 \log_{10}\Big(\frac{kT}{\rm 1 \ keV}\Big) 
\end{equation}
\begin{equation} \label{eqn:mbhmjoint}
 \log_{10}\Big(\frac{M_{\rm BH}}{10^9M_{\rm \odot}}\Big) = -1.07 + 1.94 \log_{10}\Big(\frac{M_{\rm bulge}}{10^{11}M_{\odot}}\Big) 
\end{equation}
\begin{equation} \label{eqn:mbhcmjoint}
 \log_{10} \Big(\frac{M_{\rm BH}}{10^9M_{\rm \odot}}\Big) = -0.82 + 1.16 \log_{10} \Big(\frac{M_{\rm 500}}{10^{13} M_{\rm \odot}}\Big) 
\end{equation}
\begin{equation} \label{eqn:mbulget}
\log_{10} \Big(\frac{M_{\rm Bulge}}{10^{11}M_{\rm \odot}}\Big) = 0.48 + 0.92 \log_{10}\Big(\frac{kT}{\rm 1 \ keV}\Big)
\end{equation}

The intrinsic scatter in both axis was also calculated for the joint sample using Equation \ref{eqn:intrinscatter}, as well as the Pearson and Spearman correlation coefficients. All results are summarized in Table \ref{tab:paramsjoint}. The joint fit relations are visualized in Figure \ref{fig:scatt} as the dotted green line. From Table \ref{tab:paramsjoint}, it is clear that the \MbhT relation is slightly tighter than the \MbhM relation. This is likely due to the relation for the joint fit being driven by the dynamically measured BHs. The scatter in both the \MbhT and the \MbhM relation has been reduced, however the scatter arising from the Fundamental Plane relation likely plays a notable role. 

\begin{figure*}
\centering
\epsscale{1.5}
\includegraphics[width=0.96\textwidth]{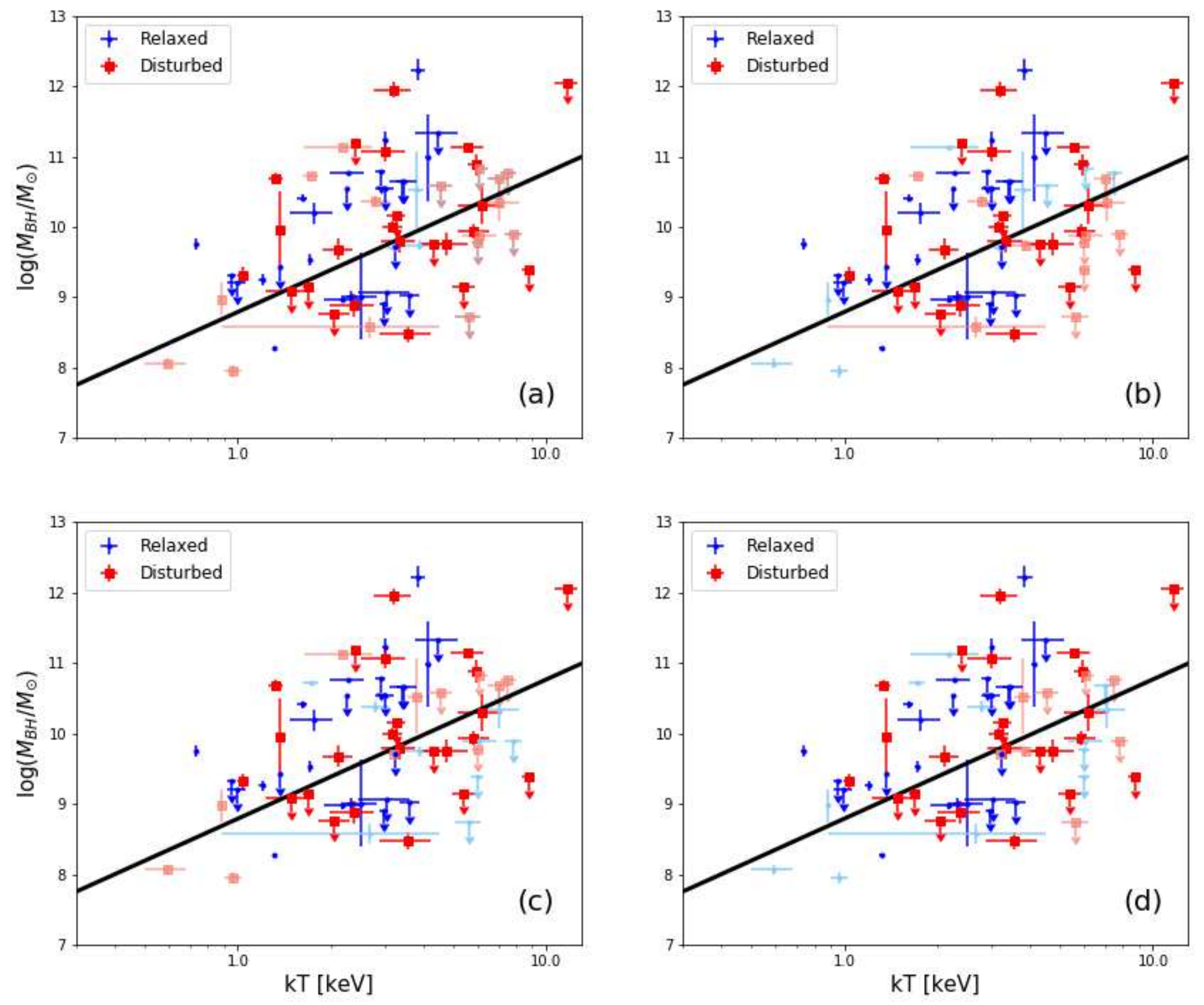}
\caption{Plot of BH mass against cluster temperature for each of the four parameters used to distinguish between relaxed and disturbed systems; (a) Central Electron Density, (b) Central Entropy, (c) Cuspiness and (d) Concentrated Surface Brightness. The blue circles and red squares indicate relaxed and disturbed clusters respectively where points represented with arrows are upper limits. Darker colours indicate sources which were identified as either CC-like or NCC-like by more than three parameters. The solid line represents the best-fit \MbhT relation obtained for the full joint sample (Equation \ref{eqn:mbhtjoint}).} \label{fig:mbhtcc}
\end{figure*}

\subsection{Monte Carlo simulations}
\label{sec:mc}
Given that the scatter in the \MbhT and the Fundamental Plane relation are comparable, we investigate whether the intrinsic scatter of the Fundamental Plane relation may be responsible for the observed scatter in the the  \MbhT relation. Therefore, we performed Monte Carlo simulations. 

We assumed that the relation in Equation \ref{eqn:mbht} was perfect and used it to determine the masses of BHs using a random distribution of $10^4$ temperatures within the temperature range of our data. A Gaussian distribution was created using a scatter of 0.88 that characterizes the Fundamental Plane relationship of \citet{merloni03}, which was convolved with the BH masses produced to give a randomly scattered sample of simulated data. We then calculated the best-fit relation using BCES\textunderscore REGRESS code and the intrinsic scatter in the y-direction ($\sigma_{\rm intrin}^y$) for the simulated data. We repeated this $10^3$ times, and derived the gradient, y-intercept, and $\sigma_{\rm intrin}^y$ for each realization.

The left panel of Figure \ref{fig:simulation} shows one representative realization of the simulation along with the observed data for the \MbhT relationship. All the observed points lie within the range set by the simulated data, hinting that at least part of the scatter in our data may be due to the Fundamental Plane relation. 

In the right panel of Figure \ref{fig:simulation} we show the distribution of $\sigma_{\rm intrin}^y$ for the  simulations. We find that $68\%$ of the simulation have a scatter in the range of $\sigma_{\rm intrin}^y = 0.87-0.89$, which is similar the scatter of the Fundamental Plane  relation. We find that the observed scatter of $\sigma_{\rm intrin}^y = 0.91$ is observed in none of the realizations. It is possible that dominant fraction of the scatter is introduced due to the Fundamental Plane relation that was used to infer the BH masses but further factors may also contribute to the scatter. These are further discussed in Section \ref{sec:discuss}.

\subsection{Investigating cool core and non-cool core clusters}
\label{sec:ccnccresults}

In Section \ref{sec:ccncc}, we used the median of each morphological parameter to split hte sample into the most relaxed (CC like) and most disturbed (NCC like) systems.

In Figure \ref{fig:mbhtcc} we plot the BH mass against cluster temperature for each of the four parameters to probe if there is any distinction between the two types. We find that relaxed and disturbed clusters do not occupy different parts of the parameter space. This hints that BHs do not undergo drastically different evolution in the centers of these two categories of cluster. Therefore although galaxy/cluster mergers and cooling flows could be contributing to the unexpected growth of the BH in the BCG, we cannot identify the most significant cause. To further investigate the relations, we make a distinction between clusters for which three or more parameters identified them as relaxed or disturbed. However, we do not find statistically significant differences between the two populations. We implemented the same fitting method as in \ref{sec:correlations} using the BCES\textunderscore REGRESS code for each of the parameters. We found that for each parameter, the difference in the \MbhT relation for the two sub-samples was statistically insignificant, therefore we still cannot justify the cause of the unexpected growth for clusters that resemble properties similar to CC and NCC systems.




\section{Discussion} \label{sec:discuss}
\subsection{Tightness of the scaling relations} 
\label{sec:tightness}

In this work, we investigated the tightness of the relations between the
total mass of galaxy clusters (traced by the best-fit gas temperature),
the stellar mass of BCGs, and the mass of the central BH.

We found that the best-fit \MbhT relation is similar to
that found by \citet{bogdan17}, albeit the scatter is significantly
larger (Table \ref{tab:params}). While the dominant fraction of the
scatter likely originates from the intrinsic scatter in the fundamental plane
relation, part of the scatter in the relation may arise from
uncertainties in measuring the X-ray and radio luminosity of the BH as
well as measuring the cluster temperature. Specifically, resolving the
nuclear X-ray source and deriving its luminosity is challenging in the
center of luminous galaxy clusters \citep{mezcua17}. In a
fraction of the BCG sample, the BHs remained undetected, which might be
either due to the dormant nature of the BHs or the luminous ICM in which
the BCG is embedded. Thus, the large observed scatter in the \MbhT relation is
likely due to the combination of the intrinsic scatter of the Fundamental Plane 
relation, measurement uncertainties in the X-ray luminosity of
the BH and ICM temperature, and the intrinsic scatter in the relation.

The scatter in the \MbhM relation is comparable to that obtained for the
\MbhT relation. Because the scatter for both relations is dominated by
the intrinsic scatter of the Fundamental Plane, our results do not contradict
\citet{bogdan17}, who found lower scatter and tighter correlation for
the \MbhT relation. 
This result is further supported by the joint fitting of this  and \citet{bogdan17} data sets. Again the \MbhT relation is tighter than the $M_{\rm BH} -M_{\rm bulge}$, but both exhibit similar scatter.

The discrepancies in the \MbhT relation between our work and that of \citet{bogdan17}  are likely due to the different populations of galaxy groups and clusters. The sample of \citet{bogdan17} was mainly dominated by galaxy groups, and their sample included only two systems with an ICM temperature greater than 2 keV. This implies that their relation in the high mass end is not well constrained. In this work, we mainly study massive galaxy clusters. Indeed this sample consists of $53$ clusters with a temperature greater than $2$ keV. Therefore the nature of the \MbhT relationship is dictated largely by the high mass clusters resulting in a steeper relation. 


Our results suggest that the \MbT is the tightest relationship (Table
\ref{tab:params}). The scatter and the tightness of this relation is
comparable with that obtained in \citet{bogdan17}. However, we note that
both $M_{\rm bulge}$ and $kT$ are directly measured quantities, unlike
those where the $M_{\rm BH}$ is inferred from the Fundamental Plane. Therefore, it is
not surprising that there is less scatter observed in this relationship. From the joint fitting, we can draw a similar conclusion. 

Hence, the observed scaling relations are not inconsistent with the
results of \citet{bogdan17}. However, to conclusively confirm that the \MbhT
relation is tighter than the \MbhM relation, more accurate BH mass
measurements would be required.

\subsection{Processes aiding the growth of BHs in BCGS}
\label{sec:processes}

Our results demonstrate that for a given stellar bulge mass, BCGs have more massive BHs than satellite galaxies. This can be seen in panel (b) of Figure \ref{fig:scatt} where we have plotted BH mass against BCG stellar mass. On this plot we added the \MbhM relation of \citet{mccon13} to emphasize how overly massive most of these particular BHs are.  It is worth noting that the best-fit \MbhM relation presented in \citet{mccon13} (see Figure \ref{fig:scatt}) includes BCGs. Excluding BCGs would result in a shallower \MbhM relation, implying that the BHs studied in this work would be even stronger outliers. Therefore, we overview the physical processes that may aid the growth of these BHs.

A candidate process that could be contributing to the accelerated growth
of BHs in BCGs is galaxy-galaxy mergers. Mergers play an important role
in the growth of BCGs since BCGs grow a factor of  $1.8$ in mass between
z=1 and the present-day universe \citep{burk13}. At the centers of galaxy clusters,
mergers are more frequent than in field or satellite galaxies. Clusters with similar total mass can exhibit different BCG merging histories. Within 50 kpc radius a
BCG can have on average $\sim6.45$ companions, which over time will
infall and merge with the BCG \citep{burk13}. A merger of a gas-rich galaxy could directly feed the
BH and induce star formation, hence the $M_{\rm BH} - M_{\rm bulge}$
relation may still hold. However, \citet{kav14} suggested that certain
merger events weaken the coupling between stellar mass and BH growth
which would allow for a larger scatter in this relationship, especially
for BCGs.

Structure formation simulations established that BHs with masses $ > 10^9 \rm{M_{\rm \odot}}$ predominantly grow through BH-BH mergers  \citep{dubo14}. For central cluster galaxies BH-BH mergers, in which one of the BHs does not get ejected from the center, occurs more frequently than in field or satellite galaxies \citep[see][and references therein]{yoo04}]. In addition, BCGs undergo many dry mergers. During these gas-poor mergers, the star formation is less likely to be induced, but the central BH will still grow due to the BH-BH merger \citep{volon13}. These processes together could result in the higher BH masses of BCGs. Therefore it is possible that mergers aid the growth of BHs in BCGs.

Cold gas flows that directly feed the BH could also play a role in
aiding the growth of BHs. Cold gas flows may occur within a radius
occupied by the BCG \citep{reis96}, if the cooling time of the gas is
shorter than the Hubble-timescale. While subsonic cold flows may enhance
star formation \citep[see][and references therein]{odea10}, many BCGs
with cooling flows do not show signatures of active star formation. It
is possible that the low-angular momentum gas flows do not give rise to
significant star-formation, but support the rapid growth of BHs. Three
clusters exhibit cooling flows in our sample, A1795, A2597 and Hydra,
which cover a broad range of cluster temperatures with $kT = 1.72 - 6.07$
keV \citep{odea04,raff06}. The inferred mass of these BHs exceeds the
expected value from the \MbhM relation of \citet{mccon13}, these particular clusters are 
$3.5\sigma-4.0\sigma$ outliers from the local \MbhM scaling relation. In
addition, \citet{raff06} examined the star formation rate and BH growth
rate of these clusters, and concluded that for Hydra the BH is growing
faster than expected from the relation of \citet{mag98}. We also
examined the location of these BHs on the \MbhT and \MbhM relations. We
find that all three are within $1\sigma_{\rm intrin}^y$ of the \MbhT
relation. However, they are $2\sigma_{\rm intrin}^y$ above our best-fit
\MbhM relation. Overall, these results are consistent with the findings
of \citet{raff06}. Therefore, it is feasible that low angular momentum
cold flows play a notable role in fueling the growth of BHs in BCGs. This is in slight contradiction with the results found in Section \ref{sec:ccnccresults}, where we observed that CC clusters do not host significantly larger BHs than NCC ones. However, our sample only has three clusters with distinct cooling flows, therefore any dependence of the \MbhT relation on CC properties may have been washed out due to the sample splitting process.


\section{Conclusions}
\label{sec:conclusion}
In this work, we investigated the relationship between cluster mass, BCG stellar mass, and BH mass. The main results can be summarized as follows:
\begin{itemize}

\item We studied a sample of 71 galaxy clusters in the redshift range of
$z = 0.006 - 0.29$. The BH mass of the BCGs was inferred from the
Fundamental Plane relation and the total mass of the galaxy clusters
was traced through the temperature of the ICM.

\item We concluded that the BH mass of BCGs significantly exceeds that
expected from the local scaling relations, implying that additional
processes aid the growth of these BHs. We also derived scaling relations
between the BH mass, BCG stellar mass, and galaxy cluster mass.

\item We found that the best-fit \MbhT relation is steeper than that
of \citet{bogdan17} and a larger scatter is present. Using Monte
Carlo simulations we determined that most of this scatter may originate
from the intrinsic scatter of the Fundamental Plane relation.

\item We split the galaxy cluster sample using different criteria, and explored whether cool core and non-cool core clusters exhibit a
different  \MbhT relation. However, we did not find a statistically significant difference, suggesting the BHs in BCGs do not undergo different evolution on CC and NCC clusters. 

\item We discussed the potential causes of the unexpected growth in these BHs, from BH-BH mergers up to cluster-cluster mergers and cooling flows. Some clusters in our sample exhibited these features. Hence we compared their BH masses to the relations we found in section \ref{sec:correlations}, which showed that these processes may have some influence over the aided growth of BHs in BCGs. However due to the small number of objects in our sample displaying either a cluster-cluster merger or a cooling flow are small, we cannot definitively say to what extent the effect is that these processes have. We conclude that in order to test the effects of these processes, we will need to probe galaxy evolution simulations.

\end{itemize} 

\acknowledgements

We thank Vinay Kashyap and the Statistical Consulting Service of the Harard Statistics Department for helpful discussion about the fitting procedures. This research has made use of Chandra data provided by the Chandra X-ray Center (CXC). The publication makes use of software provided by the CXC in the application package CIAO. In this work the NASA/IPAC Extragalactic Database (NED) has been used. \'A.B. acknowledges support from the Smithsonian Institute.  L.L. acknowledges support from the Chandra X-ray Center through NASA contract NNX17AD83G. M.V. acknowledges funding from the European Research Council under the European Communitys Seventh Framework Programme (FP7/2007-2013 Grant Agreement no. 614199, project BLACK). 

\newpage

\appendix
\section{}
Here we display images of each cluster as seen in the X-ray as well as displaying the annulus within which the final spectra were extracted. 


\begin{figure}[H]
\centering
\includegraphics[width=0.80\textwidth]{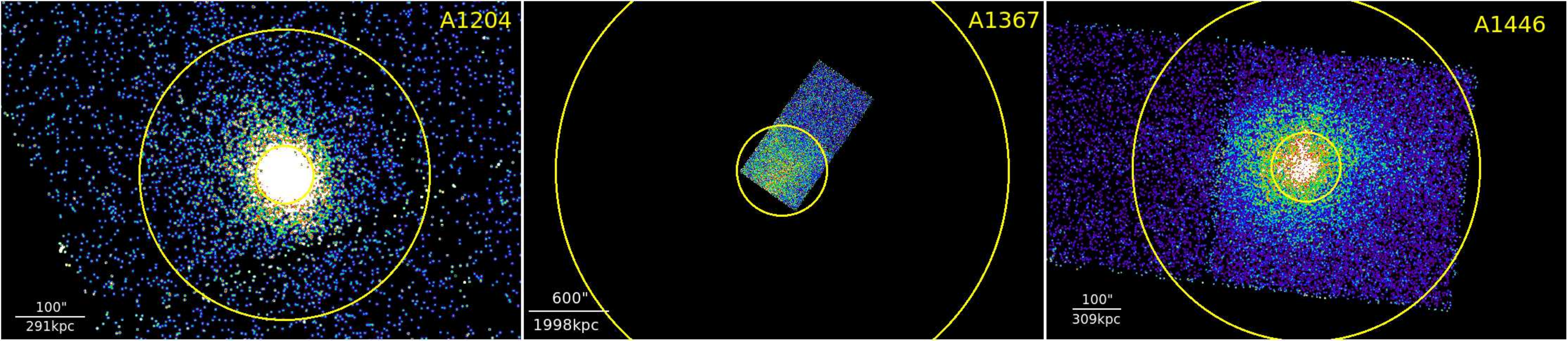}

\includegraphics[width=0.80\textwidth]{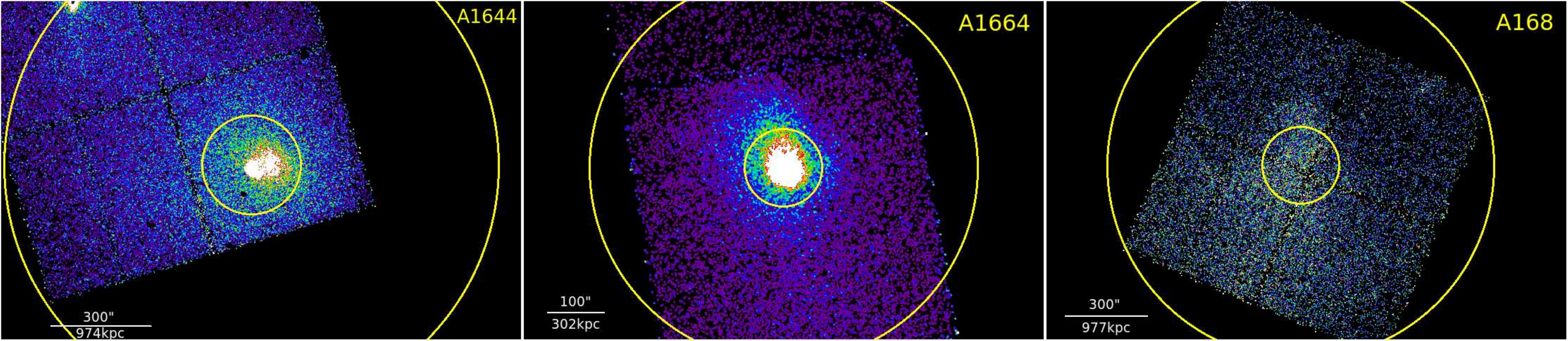}

\includegraphics[width=0.8\textwidth]{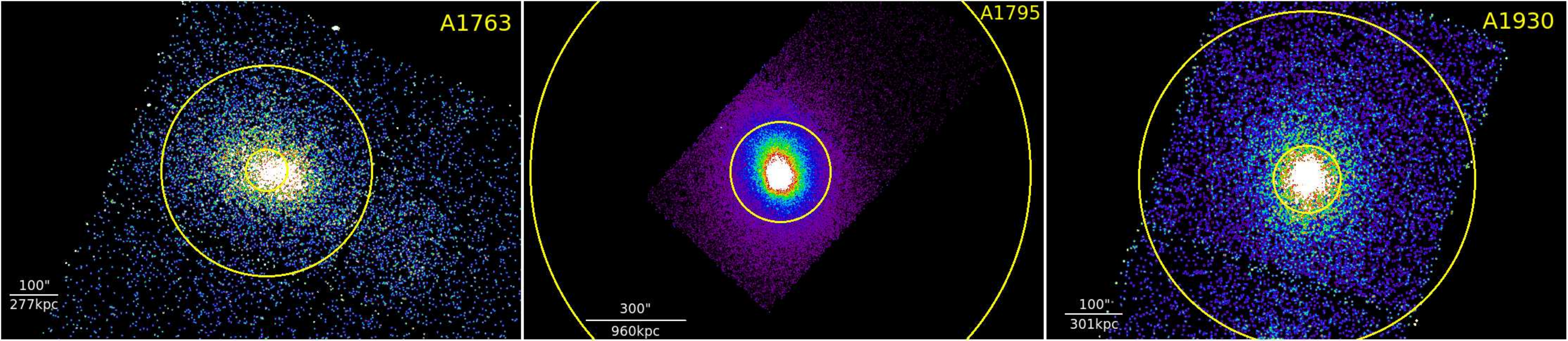}

\includegraphics[width=0.8\textwidth]{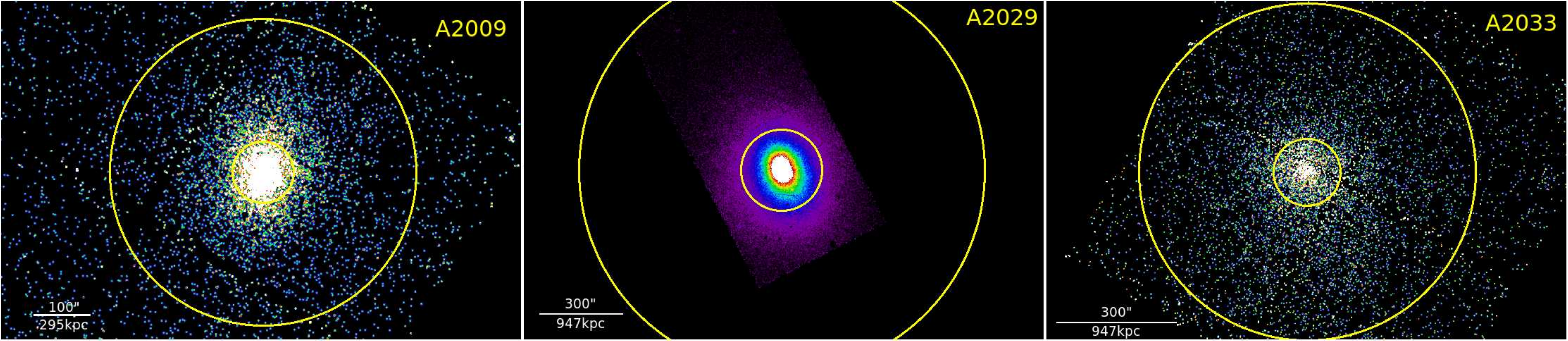}

\caption{} \label{fig:tile1}
\end{figure}

\begin{figure*}
\centering
\includegraphics[width=0.8\textwidth]{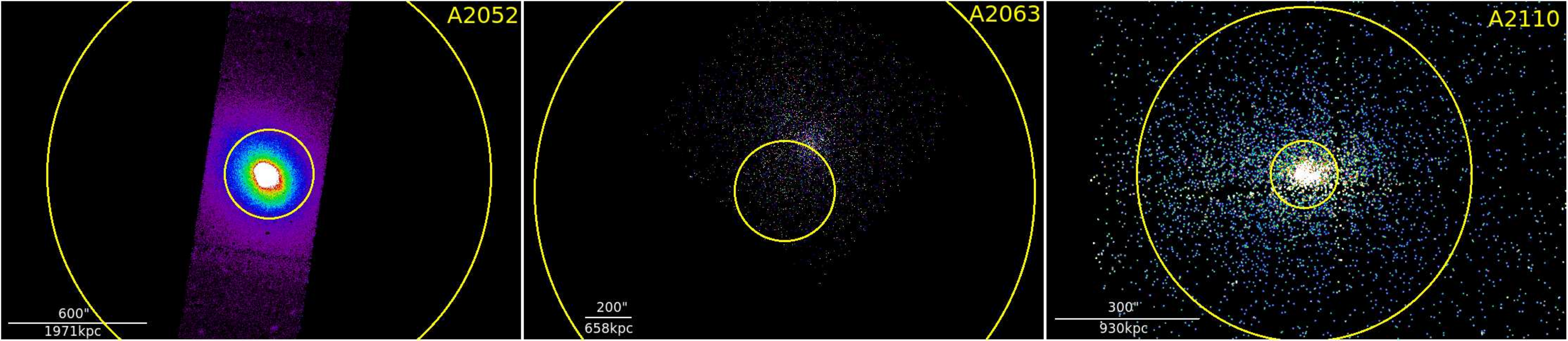}

\includegraphics[width=0.8\textwidth]{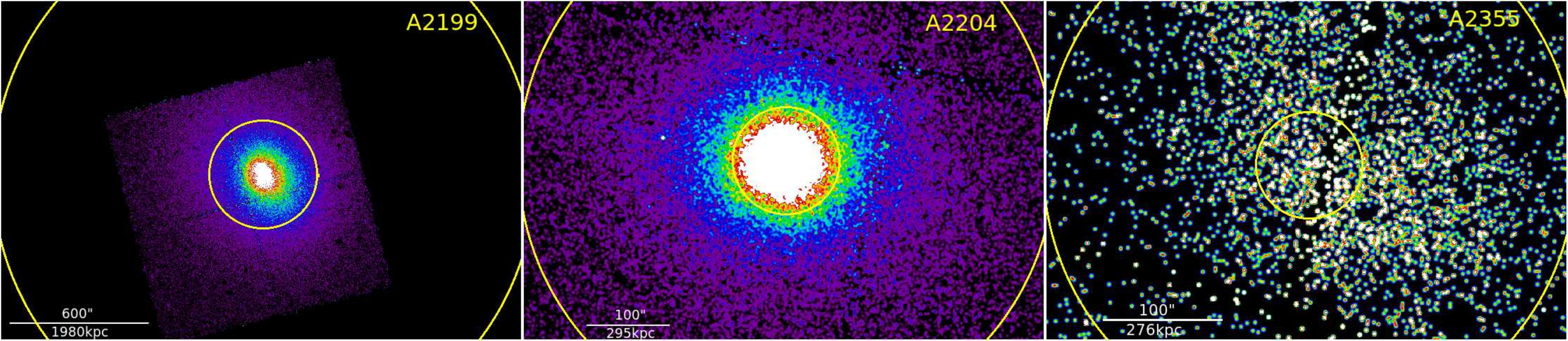}

\includegraphics[width=0.8\textwidth]{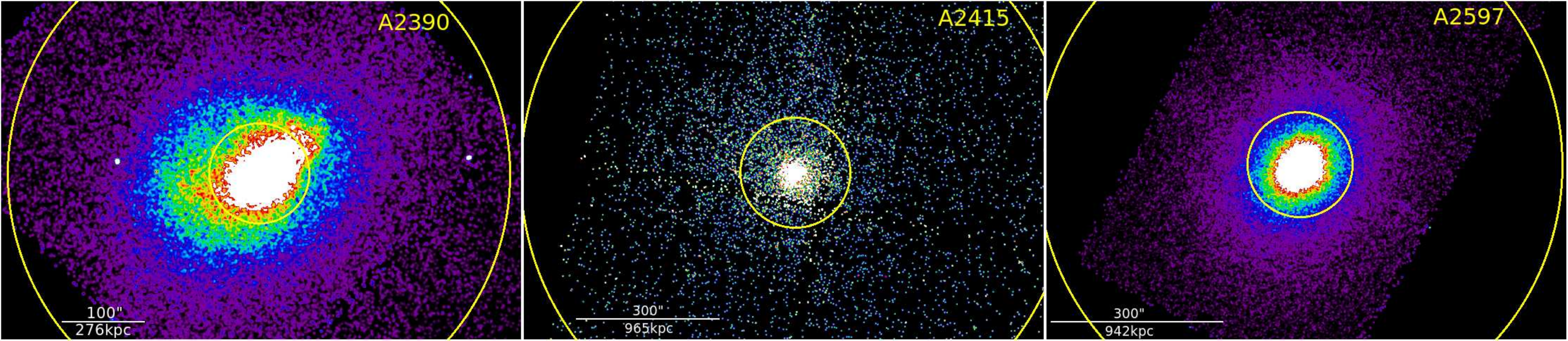}

\includegraphics[width=0.8\textwidth]{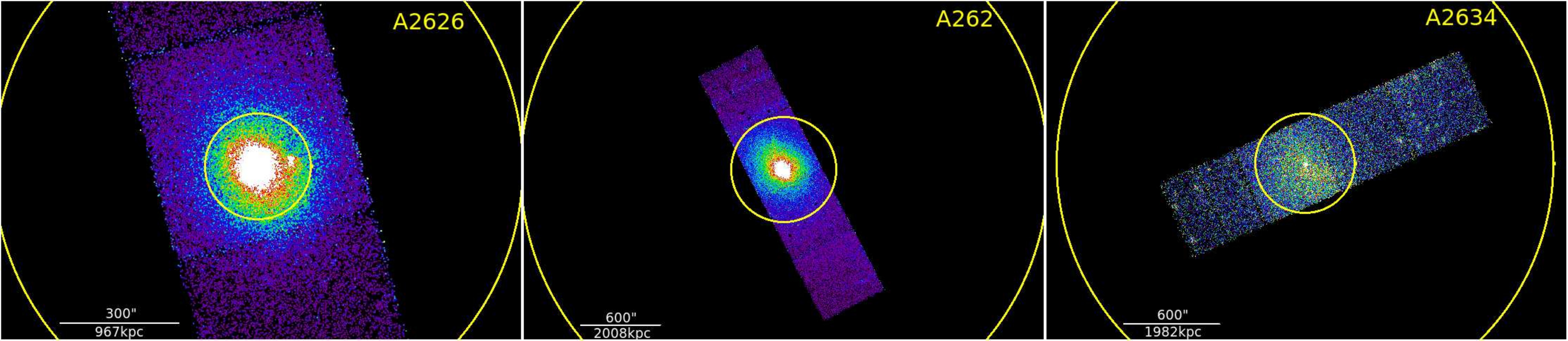}

\includegraphics[width=0.8\textwidth]{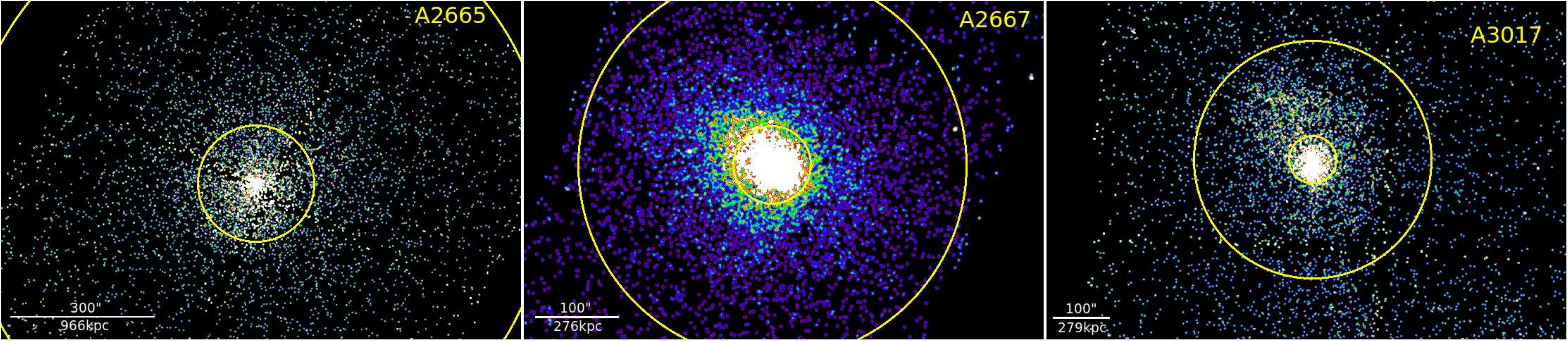}
\caption{} \label{fig:tile2}
\end{figure*}
\begin{figure*}
\centering
\includegraphics[width=0.8\textwidth]{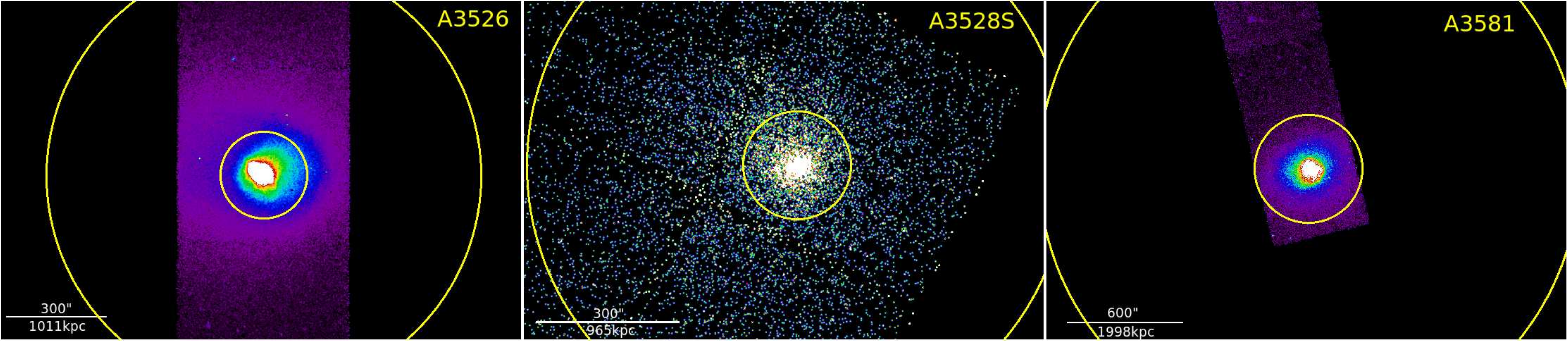}

\includegraphics[width=0.8\textwidth]{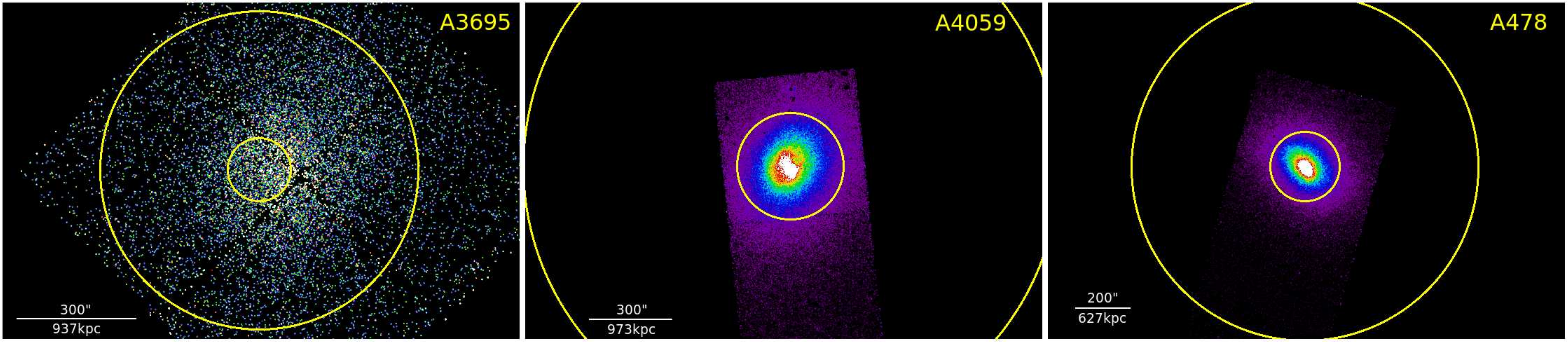}

\includegraphics[width=0.8\textwidth]{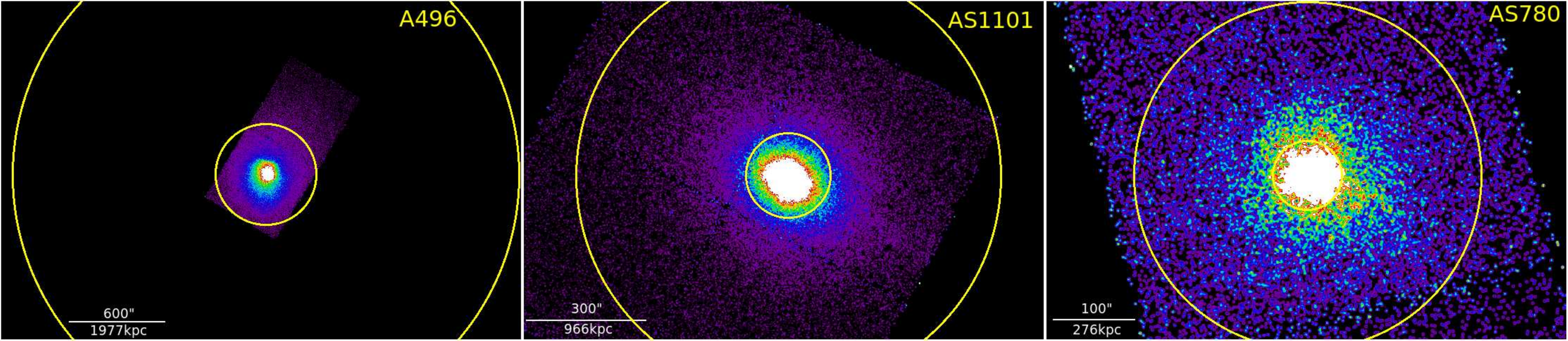}

\includegraphics[width=0.8\textwidth]{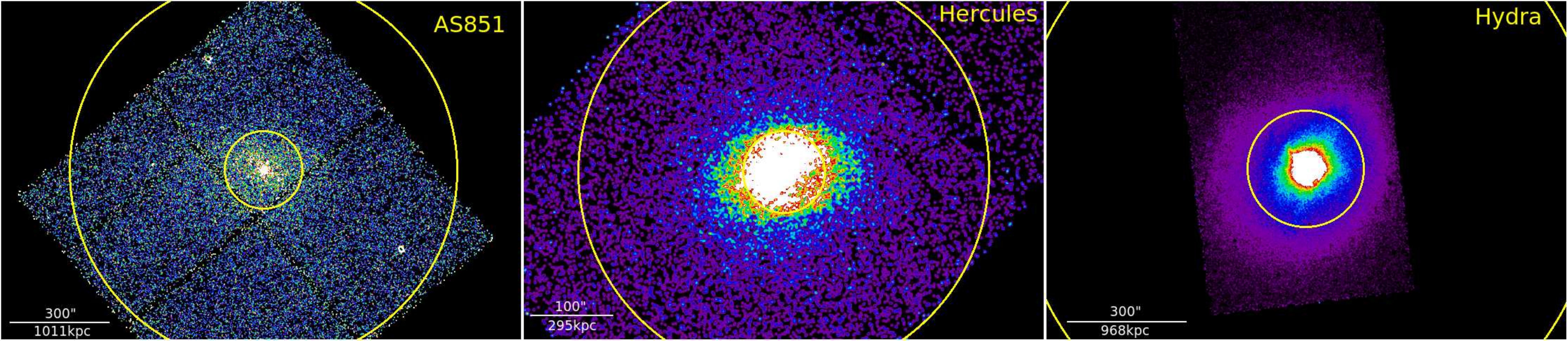}

\includegraphics[width=0.8\textwidth]{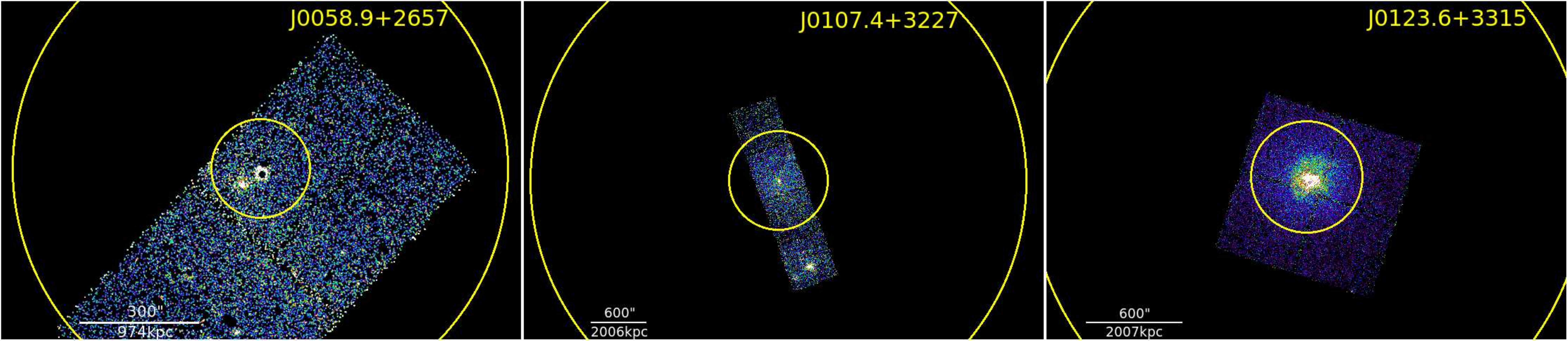}
\caption{} \label{fig:tile3}
\end{figure*}
\begin{figure*}
\centering
\includegraphics[width=0.8\textwidth]{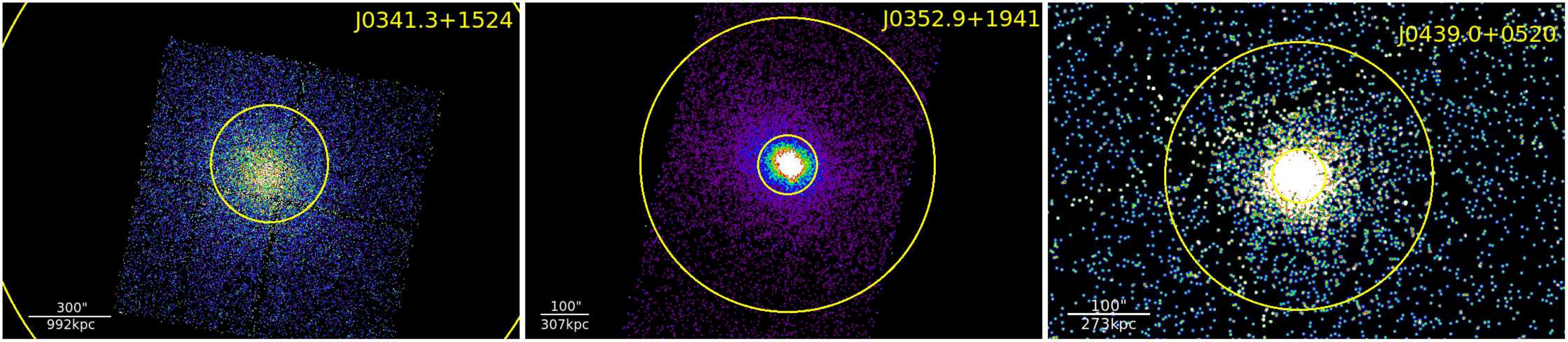}

\includegraphics[width=0.8\textwidth]{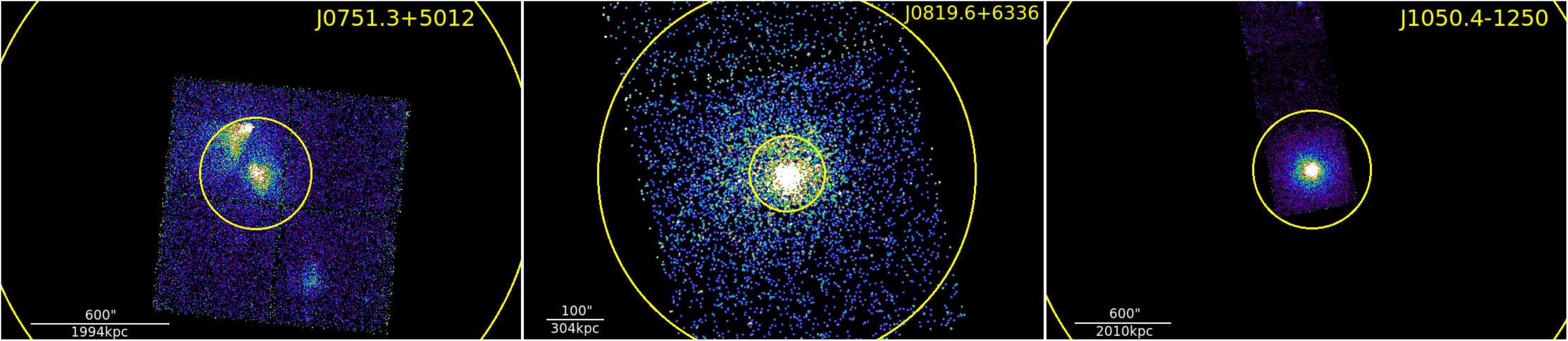}

\includegraphics[width=0.8\textwidth]{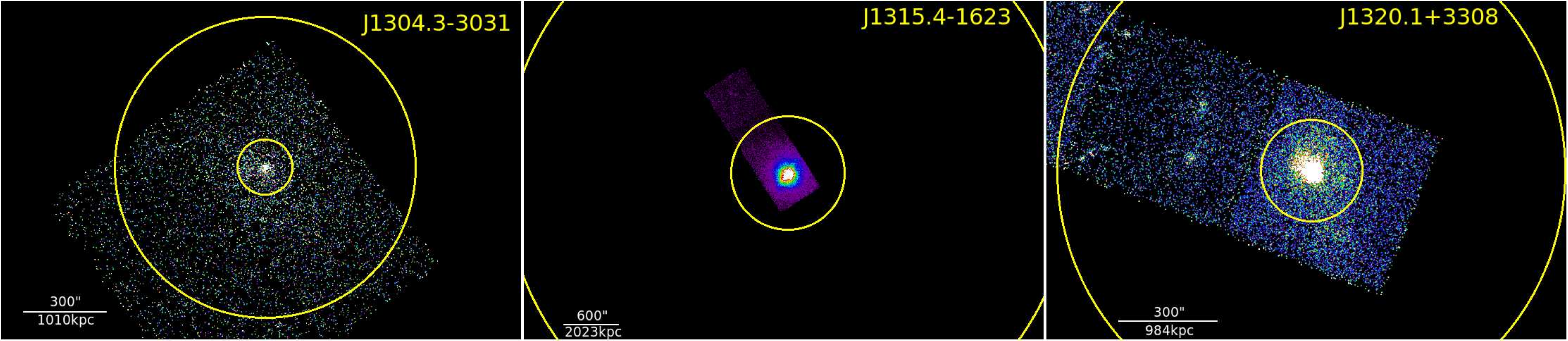}

\includegraphics[width=0.8\textwidth]{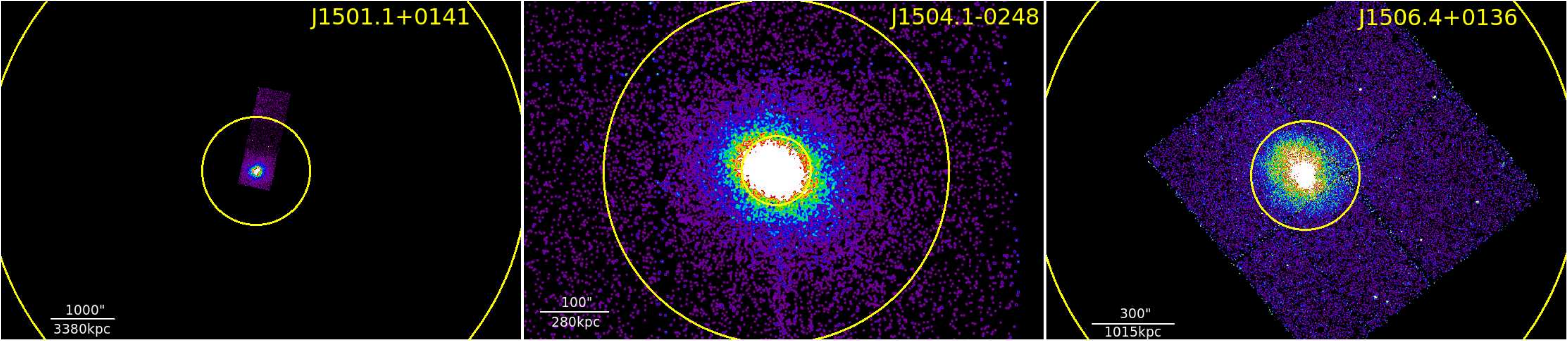}

\includegraphics[width=0.8\textwidth]{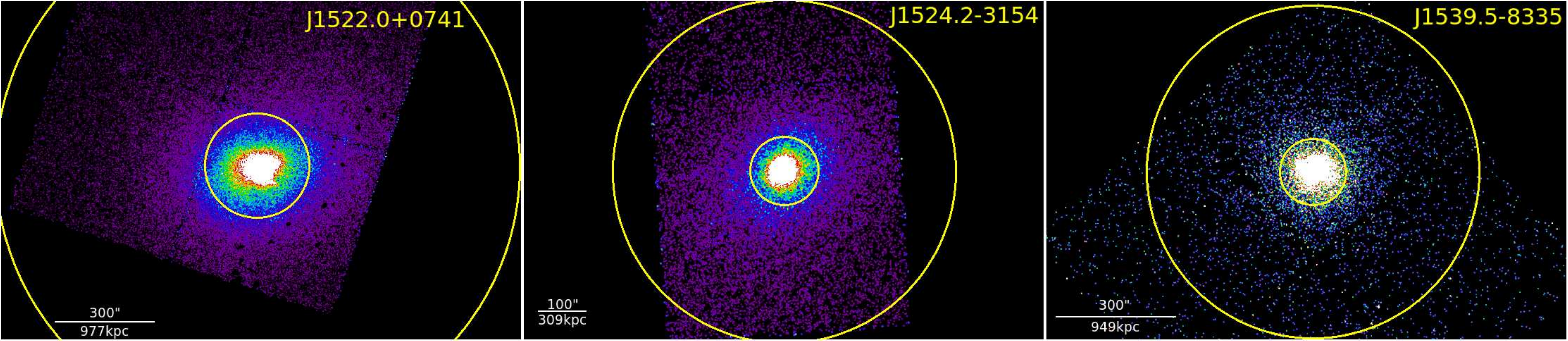}
\caption{} \label{fig:tile4}
\end{figure*}
\begin{figure*}
\centering
\includegraphics[width=0.8\textwidth]{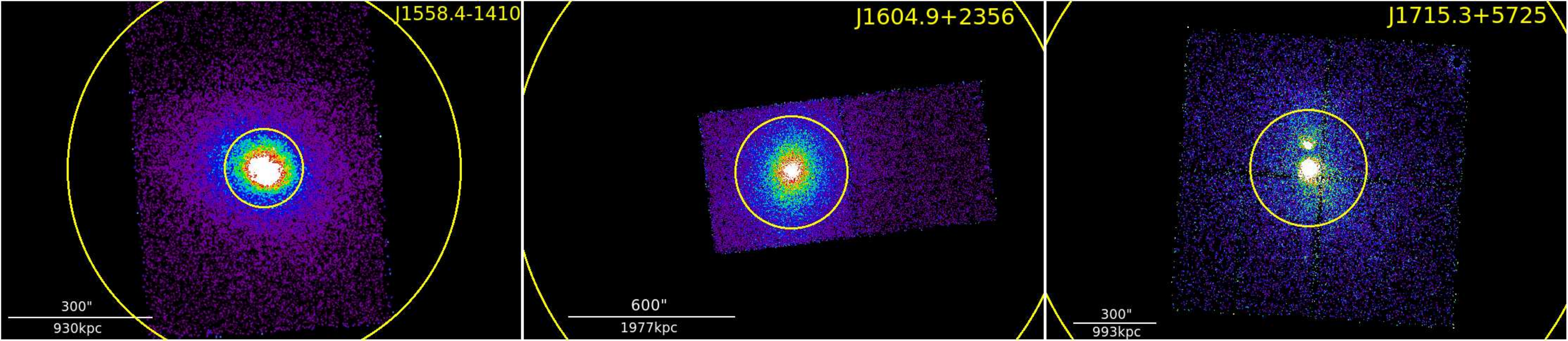}

\includegraphics[width=0.8\textwidth]{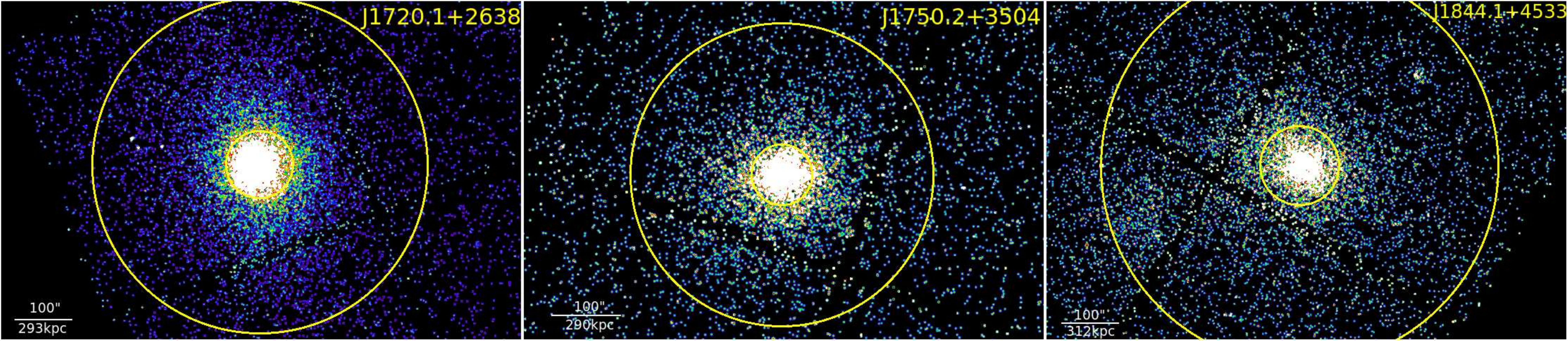}

\includegraphics[width=0.8\textwidth]{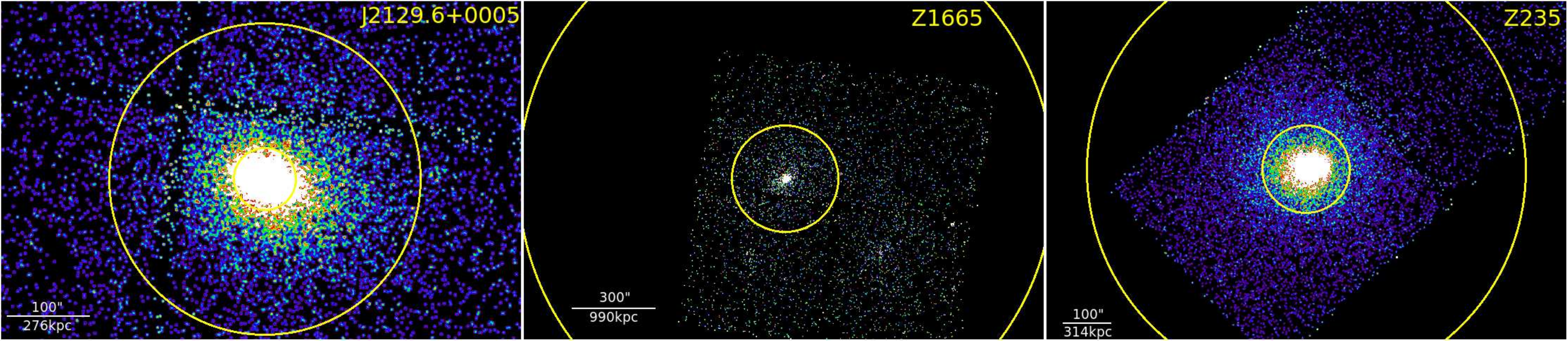}

\includegraphics[width=0.8\textwidth]{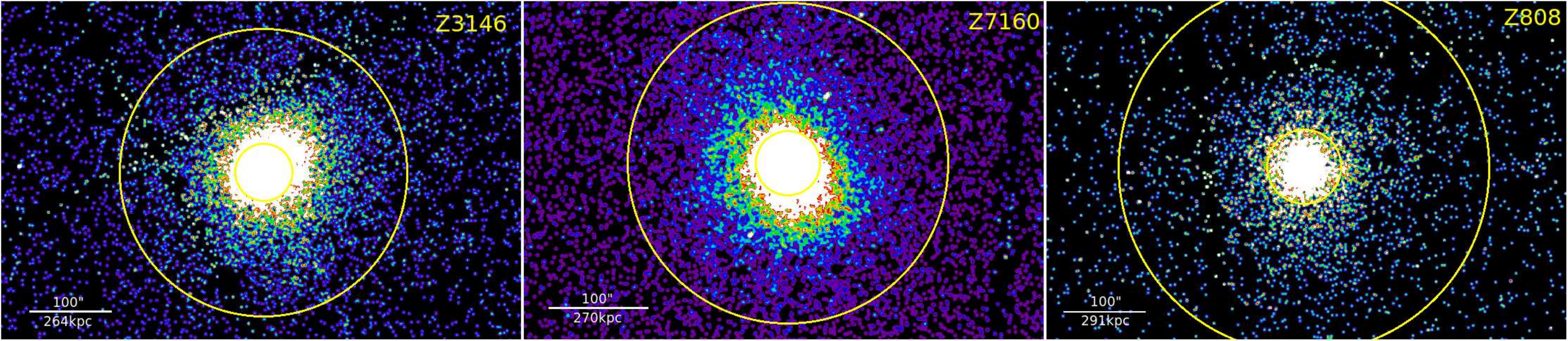}

\includegraphics[width=0.8\textwidth]{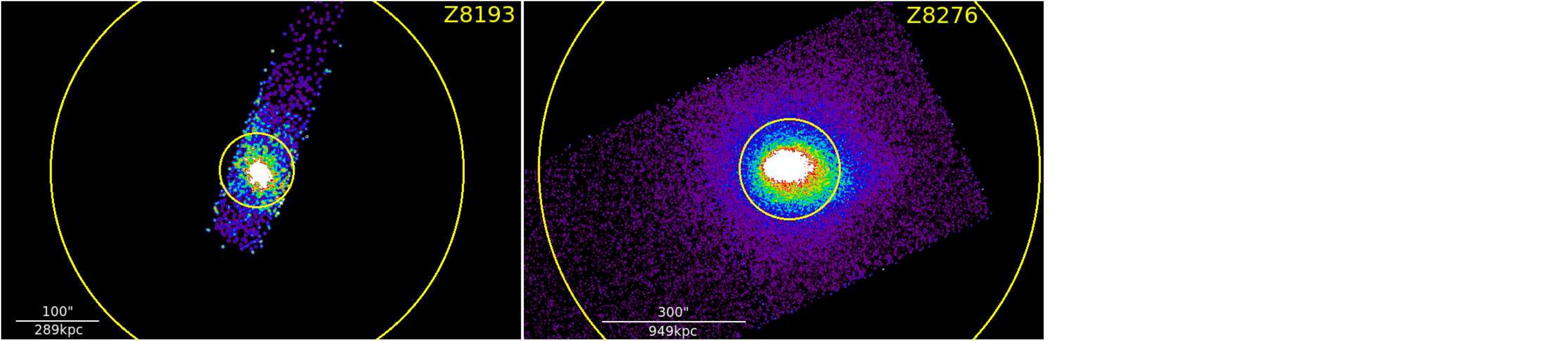}
\caption{} \label{fig:tile5}
\end{figure*}

\end{document}